\documentstyle[12pt,amssymb,amscd]{article}   

\def\AFOUR{%
\setlength{\textheight}{9.0in}%
\setlength{\textwidth}{5.75in}%
\setlength{\topmargin}{-0.375in}%
\hoffset=-.5in%
\renewcommand{\baselinestretch}{1.17}%
\setlength{\parskip}{6pt plus 2pt}%
}
\AFOUR                                           
\def\car{\mathop{\square}}
\def\carre#1#2{\raise 2pt\hbox{$\scriptstyle #1$}\car_{#2}}

\makeatletter
\def\section{\@startsection {section}{1}{\z@}{-3.5ex plus -1ex minus
 -.2ex}{2.3ex plus .2ex}{\large\bf}}
\def\subsection{\@startsection{subsection}{2}{\z@}{-3.25ex plus -1ex minus
 -.2ex}{1.5ex plus .2ex}{\normalsize\bf}}
\makeatother
\makeatletter
\@addtoreset{equation}{section}
\renewcommand{\theequation}{\thesection.\arabic{equation}}
\makeatother
\newcommand{\nc}{\newcommand}
\newcommand{\rnc}{\renewcommand}
\nc{\be}{\begin{equation}}
\nc{\ee}{\end{equation}}
\nc{\bea}{\begin{eqnarray}}
\nc{\eea}{\end{eqnarray}}

\def\slash#1{\setbox0=\hbox{$#1$}#1\hskip-\wd0\hbox to\wd0{\hss\sl/\/\hss}}

\def\href#1#2{{#2}}

\rnc{\a}{\alpha}
\nc{\ab}{\bar{\a}}
\nc{\ap}{\a^{+}}
\nc{\abm}{\ab^{-}}
\rnc{\b}{\beta}
\nc{\bb}{\bar{\b}}
\nc{\bbp}{\bb_{\zb}^{+}}
\nc{\bm}{\b_{z}^{-}}
\nc{\oa}{\overline{\a}}
\nc{\ob}{\overline{\b}}
\rnc{\gg}{\gamma}
\rnc{\d}{\delta}
\nc{\f}{\phi}
\nc{\fb}{\bar{\phi}}
\nc{\vf}{\varphi}
\nc{\p}{\psi}

\rnc{\c}{\chi}
\nc{\la}{\lambda}
\nc{\m}{{\mathrm m}}
\nc{\n}{\nu}
\rnc{\o}{\omega}
\nc{\Om}{\Omega}
\rnc{\t}{\theta}
\nc{\eps}{\epsilon}
\rnc{\S}{\Sigma}
\nc{\F}{\Phi}
\nc{\trac}[2]{{\textstyle\frac{#1}{#2}}}
\nc{\ex}[1]{\mbox{e}^{\,\textstyle#1}}
\nc{\mat}[4]{\left(\begin{array}{cc}#1&#2\\#3&#4\end{array}\right)}
\nc{\som}[9]{\left(\begin{array}{ccc}#1&#2&#3\\#4&#5&#6\\#7&#8&#9%
\end{array}\right)}
\nc{\tr}{\mathop{\mbox{tr}}\nolimits}
\nc{\ad}{\mathop{\mbox{ad}}\nolimits}
\nc{\Tr}{\mathop{\mbox{Tr}}\nolimits}
\nc{\Det}{\mathop{\mbox{Det}}\nolimits}
\nc{\rk}{\mathop{\mbox{rk}}\nolimits}
\nc{\ra}{\rightarrow}
\nc{\Ra}{\Rightarrow}
\nc{\LRa}{\Leftrightarrow}
\nc{\ot}{\otimes}
\rnc{\ss}{\subset}
\nc{\nul}{\noindent\underline}
\nc{\non}{\nonumber\\}
\nc{\subs}[1]{{\vspace*{0.5cm}}%
{\noindent\underline{#1}}{\addcontentsline{toc}{subsection}{#1}}%
{\vspace*{0.3cm}}}
\nc{\zb}{\bar{z}}
\rnc{\lg}{\frak{g}}
\nc{\lt}{\frak{t}}
\nc{\lk}{\frak{k}}
\nc{\lh}{\frak{h}}
\nc{\pik}{\Pi_{\lk}}
\nc{\pip}{\Pi_{+}}
\nc{\pim}{\Pi_{-}}
\nc{\pih}{\Pi_{\lh}}
\nc{\jz}{J_{z}}
\nc{\jzh}{\jz^{\lh}}
\nc{\jzp}{\jz^{+}}
\nc{\jzm}{\jz^{-}}
\nc{\del}{\partial}
\nc{\dz}{\del_{z}}
\nc{\dzb}{\del_{\bar{z}}}
\nc{\az}{A_{z}}
\nc{\azb}{A_{\bar{z}}}
\nc{\g}{g^{-1}}
\nc{\dw}{\Delta_{W}}
\nc{\Ad}{{\mbox{Ad}}}
\nc{\ks}{Ka\-za\-ma-\-Su\-zu\-ki}
\nc{\KS}{\ks}
\nc{\ksm}{\ks\ model}
\rnc{\AA}{{\Bbb A}}
\nc{\BB}{{\Bbb B}}
\nc{\CC}{{\Bbb C}}
\nc{\PP}{{\Bbb P}}
\nc{\cpm}{\CC\PP(m)}
\nc{\cpn}{\CC\PP(n)}
\nc{\cp}[1]{\CC\PP(#1)}
\nc{\gmn}{G(m,m+n)}
\nc{\gmnk}{\gmn_{k}}
\nc{\cO}{{\cal O}}
\nc{\bcO}{\bar{\cO}}
\nc{\bO}{\bar{O}}
\nc{\oQ}{\overline{Q}}
\nc{\ie}{{\it i.e.~}}
\nc{\eg}{{\it e.g.~}}
\begin{document}
\global\parskip=4pt
\makeatother\begin{titlepage}
\begin{flushright}
{}
\end{flushright}
\vspace*{0.1in}
\begin{center}
{\Large\bf D1/D5 systems in ${\cal N}=4$ string theories} \\ 
\vskip .3in
\makeatletter




\centerline{ Edi Gava\footnote{E-mail: gava@he.sissa.it}, \ 
Amine B. Hammou\footnote{E-mail: amine@sissa.it}, \ Jose F. Morales
\footnote{E-mail: morales@phys.uu.nl }, \
{\it and}
\ Kumar S. Narain \footnote{E-mail: narain@ictp.trieste.it} } 
\bigskip 
\centerline{\it SISSA, Trieste, Italy$^{1,2}$} 

\smallskip \centerline{\it The Abdus Salam ICTP, Trieste, Italy$^{1,4}$}

\smallskip \centerline{\it INFN, Sez. di Trieste,
Trieste, Italy$^{1,2}$} 

\smallskip \centerline{\it Spinoza Institute, Utrecht, The Netherlands $^3$} 

                                                                   
\end{center}

\vskip .10in
\begin{abstract}
\noindent 
We propose CFT descriptions 
of the D1/D5 system in a class of freely 
acting $Z_2$ orbifolds/orientifolds
of type IIB theory, with sixteen unbroken supercharges. 
The CFTs describing D1/D5 systems involve 
${\cal N}=(4,4)$ or ${\cal N}=(4,0)$ sigma models on 
$(R^3\times S^1\times T^4\times (T^4)^N/S_N)/Z_2$,
where the action of  $Z_2$ is diagonal and
its precise nature depends on the model.
We also discuss D1(D5)-brane states 
carrying non-trivial Kaluza-Klein charges,
which correspond to excitations  of 
two-dimensional CFTs of the type  
$(R^3\times S^1\times T^4)^N/S_N\ltimes Z_2^N$. The resulting 
multiplicities
for two-charge bound states are shown to
agree with the predictions of U-duality.
We raise a puzzle concerning the 
multiplicities of three-charge systems, which is generically present 
in all vacuum configurations with sixteen unbroken supercharges
studied so far, including the more familiar type IIB on $K3$ case:
the constraints put on BPS counting formulae
by U-duality are apparently in contradiction with
any CFT interpretation. We argue that the presence of RR
backgrounds
appearing in the symmetric product CFT may provide a  resolution
of this puzzle. Finally, we show that the whole tower of 
D-instanton
corrections to certain ``BPS saturated couplings'' in the low
energy effective actions match with the corresponding one-loop 
threshold corrections on the dual fundamental string side. 
~

\end{abstract}

\vfill


\makeatother
\end{titlepage}
\begin{small}
\end{small}

\setcounter{footnote}{0}

\tableofcontents

\section{Introduction}
The D1/D5 system has received much attention in the 
last few years, especially for its relation 
with the physics of 5-dimensional Black Holes \cite{SV,M1,HW,DMW},
and, more recently, in the context of AdS/CFT correspondence \cite{M2,GKS}.
The cases which have been most considered in the literature
are those of D1/D5 systems
in type IIB string theory on ${\cal M}$ with ${\cal M}$ being $T^4$ or $K3$,
where $Q_5$ D5-branes are wrapped on 
${\cal M}$
and $Q_1$ D1-branes are parallel to the D5-branes and localized on 
${\cal M}$. 
In both cases the effective 
field theory describing the system is expected to flow
in the infrared limit to a ${\cal N}=(4,4)$ CFT, whose non-trivial
part is a sigma model with target the moduli space
of $U(Q_5)$ instantons on ${\cal M}$ with instanton number
$Q_1$. This space is a smooth resolution of the 
(singular) symmetric product 
${\cal M}^N/S_N$ with $N=Q_1Q_5$ and $N=Q_5(Q_1-Q_5)+1$\footnote{
The physical reason for the shift in $Q_1$ in the $K3$ 
case was first explained in \cite{BSV}.} 
for $T^4$ and $K3$ respectively \cite{SV,V}. At the level
of CFT, the resolution of singularities is 
implemented by turning on certain
marginal deformations and therefore, if one is interested 
in topological quantities like the elliptic genus,
one may as well work at a symmetric product point
in the moduli space. 

In particular, the $T^4$ case has been 
studied in great detail 
in \cite{mms}, where a multiplicity formula for three-charge
BPS states ($Q_1$,$Q_5$ and KK momentum) preserving 
$1/8$ of the bulk supersymmetry has been derived.
The formula, valid in the primitive three-charge vector case,
was shown to agree with U-duality. In the same paper,
the multiplicity formula was found to agree, where it is supposed
to, with the counting formula obtained from the dual supergravity
description in terms of KK harmonics on $AdS_3\times S^3 \times T^4$.
 
A similar detailed analysis for the D1/D5 system in backgrounds
with $16$ supercharges (in $D<10$) 
is still missing \footnote{Not to mention the case of $16$ supercharges
in $D=10$, i.e. type
I theory, which is beyond the scope of the present work.}, 
although for the case of type IIB theory on $K3$ 
tests of the proposed symmetric product CFT have
been performed in the context of AdS/CFT 
correspondence \cite{MS,DB1,DB2,DMMV}. 
Moreover, in this case, the relation
between symmetric products of $K3$ and the moduli space
of $U(n)$ instantons on $K3$ is mathematically
well understood \cite{D1,B}. 

Purpose of the present work is to discuss systematically 
various aspects of this issue in the context
of a class of theories with 16 supercharges. These are 
obtained by orbifolding/orientifolding type IIB theory with freely
acting $Z_2$ actions, which involve shifting along
some compact direction together with the action
of $Z_2$ elements such as  the world
sheet parity $\Omega$, $(-)^{F_L}$, with $F_L$ the spacetime
left-moving fermion number or $I_4$, the reflection of the 4
coordinates of a $T^4$. A prototype of these vacua: 
type IIB/$\Omega\sigma_p$, was introduced in \cite{massimo} and
termed as ``type I theory without open strings''.

Compared to the type I case, theories 
obtained  this way  have
the simplifying feature of avoiding the presence of the open
string sector (in the case where the $Z_2$ includes the
world-sheet parity operator $\Omega$) \cite{massimo}. 
Moreover, via the adiabatic argument \cite{VW, sen}, these actions are
expected to commute with $S$-duality. Therefore the U-duality group 
is still at work, relating in a non-perturbative way
various backgrounds. In particular we can relate
the D1/D5 system in one theory to fundamental string winding
plus KK momentum states in another theory. Also, we obtain
non-trivial relations for the three-charge systems in different
backgrounds. 

We will derive CFTs for the D1/D5 systems
in the various cases and test them against the U-duality
predictions. We will find agreement for the
two-charge states but generically disagreement for the 
three-charge states. For the latter states, we will point
out a similar problem in the case of type IIB theory on $K3$. 
We will argue that the presence of non-trivial RR backgrounds
in the symmetric product CFT may provide a resolution
of this problem.

The paper is organized as follows: in section 2 we will
describe the various type IIB orbifold/orientifold
backgrounds and the U-duality relations among them 
that will be relevant to
our subsequent analysis. The discussion here will 
be general for the case where the shift is transverse
to the D-branes, but we will partially cover the
logitudinal shift case too. We also discuss the role
of RR background in the U-duality relations.

In section 3 we will derive the effective field theory
for systems of D1-branes, D5-branes and D1/D5-branes
and show that they involve symmetric products of $T^4$
and $R^3\times S^1$ factors, with appropriate $Z_2$ orbifold actions, 
depending on the background in consideration. For theories
which involve orientifolding, the resulting D1/D5 CFT is
of type $(4,0)$.
In section 4 we will derive the relevant elliptic genus formulae 
for symmetric products involving both  
even and odd fields, with respect to the above extra $Z_2$ orbifold
actions. In section 5 these results will be used to show
that the  predictions of the proposed CFTs 
agree with those of
the perturbative string partition functions of the
U dual theories, for all two-charge cases, including
the D1-D5 case. 
We also point out a problem concerning the three-charge
states (D1-D5-KK), which arises in the models we are considering
and also in the more familiar $K3$ case: 
there is an apparent clash between  U-duality  
and CFT interpretation of multiplicity formulae.

In section 6 we compute the moduli dependence
of low energy couplings involving the gauge fields
arising from KK reduction in various backgrounds. We verify
the matching of one-loop expressions on the fundamental
string side with D-instanton contributions on the dual side.

In section 7, we will make some conclusions and also
comment about the puzzle concerning three-charge states
discussed in section 5. 
In appendix A we present a systematic description of the open string theory
living on intersecting D1-D5-branes, in the presence of longitudinal
shifts.
In appendix B we derive symmetric product 
partition functions for free fields acted upon by a $Z_2$ 
orbifold. This complements the more general derivation presented in
section 4. 
In appendix C we include some details of the
genus-one modular integral relevant to the computation
of low energy couplings of section 6.

\section{U-duality chain of type IIB orbifold/orientifolds}

\subsection{Transverse shift}
In this section we construct a series of five-dimensional U-dual models
with sixteen unbroken supercharges. We adopt Sen's fiberwise 
construction procedure \cite{sen} to generate lower dimensional dual pairs 
from the self-dual (under a subgroup of the full five-dimensional 
U-duality group that we will still call U) type IIB theory on $T^5$,
which is taken to be in the {12345} directions.
We will further compactify on an additional $S^1$ 
of radius $R_6$ in the 6-th direction, to accompany 
various $Z_2$ orbifold/orientifold actions with a shift of order
two along $S^1$. We will restrict
ourselves to the case of a geometrical
shift by half winding, denoted by $\sigma_{p_i}$
with $i=1,6$ according to whether the shift is longitudinal
or transverse to the brane system.
In the transverse case, which we will consider first, this results in
a factor $(-)^{p_6}$ in the untwisted sector lattice sum, $p_6$
being the momentum in the $X_6$ direction.  
By the adiabatic argument \cite{VW}, the free 
nature of these $Z_2$'s makes the orbifold 
action commuting  with S-duality.

We start by defining a U-duality chain that maps into each other
the various charges in the perturbative and solitonic spectrum 
of the toroidal type IIB parent theory. Under these duality 
transformations, perturbative symmetries of the underlying theory
such as $\Omega$ (worldsheet parity operator), $(-)^{F_L}$ (left
moving spacetime fermion number) and $I_4$ (reflection in the (2345)
plane) are mapped into each other. A prototype of such a duality chain
is displayed in table 1.1: 

$$
\begin{CD}
{\bf A}@>S>>    {\bf B}@>T_{2345}>>{\bf  C}   @>S>>  {\bf  D}  \\
NS_{12345} && D_{12345} &&    D_1      &&  F_1 \\
F_1        &&   D_1     &&    D_{12345}&& NS_{12345} \\
p_1        &&     p_1   &&       p_1   &&  p_1 \\
(-)^{F_L}I_4&& \Omega I_4 &&  \Omega    &&  (-)^{F_L} \\ 
I_4         &&     I_4  &&    I_4       && I_4  \\
(-)^{F_L}   &&  \Omega  &&  \Omega I_4  &&(-)^{F_L} I_4\\  
\end{CD}
$$\\
\hspace{2.0 cm} {\it Table 1.1: D1(D5)-p to fundamental strings}\\

For the time being different columns, labeled by {\bf A}, {\bf B},...,
represent equivalent descriptions of type IIB theory on $T^5$ but 
in the following they will stand for a triplet of models obtained
by orbifolding/orientifolding the toroidal theory by one of the three
perturbative symmetries displayed in each column 
(accompanied by a $Z_2$ shift $\sigma_{p_6}$). 
Different columns are connected by $S$ or 
$T_{ijk...}$ elements (the indices indicating the direction
along which T-duality is performed) of the U-duality group. 
Winding, momentum, NS-fivebrane
and D-brane states are denoted by 
$F_i, P_i, NS_{ijklm}, D_{ijk...}$ respectively with the indices
specifying the directions along which they are oriented.
We will focus on two-charge systems which admit
always a U-dual perturbative description in terms of winding-momentum 
charges. 

A bound state of $N$ D1-strings and $k$ units of KK momentum 
$p_1$ at step {\bf C}, for example, is mapped through $S$ (step {\bf D})
to a fundamental string wrapped $N$ times on the $1^{st}$ circle and carrying
$k$ units of momentum. Similarly a D5-p bound state at {\bf C} 
is mapped at step {\bf A} to a fundamental string bound state 
$F_1-p_1$. A D1/D5
bound state, on the other hand, can be mapped again to a fundamental
string-momentum  bound state  through the more involved chain of dualities: 
     
$$
\begin{CD}
{\bf C}@>S>>{\bf D} @>T_{15}>>{\bf E}@> S >>{\bf F}@>T_{2345}>>{\bf G}@>S>>
                                                                     {\bf H} \\
D_1      &&  F_1       &&   p_1     &&  p_1     &&       p_1  &&  p_1\\
D_{12345}&& NS_{12345} && NS_{12345} &&D_{12345} &&       D_1  &&  F_1\\
p_1      &&  p_1       &&   F_1     &&  D_1     &&  D_{12345} &&NS_{12345}\\
\Omega   && (-)^{F_L} && (-)^{F_L} && \Omega  && \Omega I_4  && (-)^{F_L}I_4\\
I_4      &&  I_4       && (-)^{F_L}I_4&&\Omega I_4&&  \Omega    && (-)^{F_L}\\
\Omega I_4&&(-)^{F_L}I_4&& I_4       &&   I_4      &&     I_4   &&  I_4\\  
\end{CD}
$$\\
{\it Table 1.2: Mapping D1/D5-branes to fundamental strings }\\

We can now consider the modding out of type IIB theory on $T^4\times S^1$
by one of the three $Z_2$'s (let say at step {\bf C}) generated
by $\Omega\sigma_{p_6}$, $I_4\sigma_{p_6}$ and $\Omega I_4\sigma_{p_6}$.
We will refer to these theories as $I$, $II$ and $III$ respectively.
Accordingly, we will denote the dual descriptions at step {\bf H}
as $I_F$, $II_F$, and $III_F$ respectively. In all the cases the shift
along $X_6$ is transverse to the D1-, D5-branes which are
wrapped along $X_1,...,X_5$.
Notice that alternative
descriptions of these triplets of theories appear at different steps
in the two tables 1.1, 1.2.

The fiberwise construction procedure \cite{VW, sen} states that 
a dual pair can be defined (under certain adiabatic hypotesis)
by modding out the parent theories by two dual actions, i.e. two
symmetry elements in the same line 
in the U-duality chain above. The inclusion of the shift 
makes the adiabatic argument applicable.
Notice also that the shift $\sigma_{p_6}$ is invariant
under all the elements of the U-duality group
involved in the above chain.

From table 1.2 we see that D1/D5 bound states in column {\bf C}
are mapped to fundamental string states in column {\bf H}, where
theories $I_F$, $II_F$ and $III_F$ appear. However, we see that
exciting KK momentum on the D1/D5 system amounts to excite
NS5-brane charges in {\bf H}. 

On the other hand the duality chains above provides
also stringent constraints on the three-charge systems
(D1-D5-KK) of our three theories:

$\bullet$ by comparing column {\bf F} with column
{\bf G} we see that $D1$ and $D5$ charges
are exchanged, with theory  $II$ left invariant,
while theories  $I$ and $III$ are exchanged.

$\bullet$ by comparing column {\bf C} with column
{\bf F} we see that $D1$ and $p$ charges
are exchanged, with theory  $I$ left invariant,
while theories  $II$ and $III$ are exchanged.

$\bullet$ by comparing column {\bf B} with column
{\bf G} we see that $D5$ and $p$ charges
are exchanged, with theory  $III$ left invariant,
while theories  $I$ and $II$ are exchanged.

As we will see, these relations will put severe
constraints on the multiplicity formulae for
the three-charge systems and hence on the
effective field theory governing them.

\subsection{Longitudinal shift}
One may ask the question of what happens if the shift is
longitudinal to the D-branes, i.e. instead of
$\sigma_{p_6}$ we have  $\sigma_{p_1}$. 

One first notice that $\sigma_{p_1}$ is still invariant under
the U-duality transformations involved in the chain 1.1, but
now in going from step {\bf D} to step {\bf E} in chain 1.2, 
the half winding shift gets transformed into a half-momentum shift 
$\sigma_{F_1}$. 
Consequently the twisted sector of the theory at 
step {\bf E} contains half-integer
momentum modes localized at fixed points.    
In going from {\bf E} to {\bf F} we expect  a non-perturbative
phase $\sigma_{D_1}$ for the states carrying D1-brane charge. 
This cannot be the whole story however.
This is clear from a comparison of the perturbative spectrum of states
at step {\bf E} and {\bf F}: 
in the former case integer (untwisted states) and 
half-integer (twisted states) momentum modes come with different 
multiplicities due to the winding shift (see formulas (\ref{iiif}) below
with $F_1\rightarrow p_1$), while in the dual description the distinction
between even and odd modes would not exist, 
since the above non-perturbative phase 
leaves invariant the whole perturbative spectrum. 

An insight about the correct map can 
be gained from a careful analysis  of the fundamental string
partition function at step {\bf E}. Model  $II$
at this step is type IIB/$(-)^{F_L}I_4\sigma_{F_1}$. As we mentioned
above, a shift in $F_1$ implies that states in the twisted sector carry 
half-integer momenta and are localized at fixed points. 
Under $S$ duality to step {\bf F}, these are mapped 
to open string states living on the D5-branes 
(``twisted sector'' with respect
to  $\Omega I_4$), sitting at orientifold
5-planes at 16 fixed points. There are, to begin with, 16 pairs
of D5-branes, each pair at a fixed point, giving rise to the gauge
group $SO(2)^{16}$. However, due to
the presence of  half-integer momentum modes $p_1$, 
we conclude that 
a $Z_2$ Wilson line along the circle on $X_1$ must be turned 
on at step {\bf F}, thereby breaking
completely $SO(2)^{16}$. If we do a further T-duality along $X_1$
we then have type I' on $T^5/Z_2$ with 32 4-branes distributed
on the 32 fixed points and a completely broken gauge group.

After four T-dualities
to step {\bf G} this, together with the fact that D5-branes sit
at sixteen different fixed points of $I_4$, translates     
into a type I theory on $T^5$ with five Wilson lines turned on to
break completely the gauge group. Finally under $S$ duality,
we get a perturbative
description of the D1/D5 system of model $II$ in terms of the fundamental
heterotic string with a gauge group completely broken by Wilson lines
at step {\bf H}. We will refer to this model as model $IV$.

We will comment later on the difficulties involved
in trying to extend this analysis to the case of longitudinal shift for
models $I$ and $III$.

\subsection{$\chi = 1/2$ point}

In the following we will be considering the symmetric
product CFTs for the D1/D5 systems. 
It is generally believed that this CFT describes the
infrared limit of the D1/D5 system for $Q_5=1$  at a point in the moduli space
where the R-R scalar $\chi$ and the 4-form $C_{(4)}$ are turned
on along the
directions transverse to the D1 brane and longitudinal to the D5 brane.
Let $V_4$ denote the volume of the $T^4$ along these directions, then
this point in the moduli space is given by
\be
\chi = \frac{C_{(4)}}{Q_1}=\frac{1}{2}, ~~~~~~~ V_4=Q_1
\ee

For non-zero $\chi$ (and $C_{(4)}$) the D1/D5 system becomes a true 
bound state since
the D1 (and D5) brane gets an induced fundamental
string charge. The net induced charges for the  system of 1 D5 brane
and  $Q_1$ D1 branes remain zero and therefore it would cost some
energy for this system to split. This phenomenon can best be
understood by mapping D1/D5 system to the perturbative states by
following the chain from step {\bf C} to step {\bf H} in the table
1.2. Actually for $\chi=1/2$, the S-duality in going from step {\bf C}
to {\bf D} should be replaced by the transformation which takes the
complex coupling constant $\tau$ to
$\frac{\tau-1}{2\tau-1}$, as pointed out in \cite{wittVS}. 
This keeps $\chi =Re(\tau) =1/2$ invariant. 
Although one can still follow the chain of
dualities, the transformations on the charge vectors are more involved
than what is presented in the table. To avoid this complication, for
this subsection, we
will replace the first three steps in the above chain by the
transformation 
$T_{15}S
T_{15}R_{15}$ where $R_{15}$ is the rotation by $\pi/2$ in the 1-5
plane. It is easy to see that this transformation takes the step {\bf
C} to {\bf H} with the charges transforming as indicated in the table.
Moreover, the complex moduli $\tau = \chi +
ie^{-\phi}$
and $\tau'= C_{(4)} + i V_4 e^{-\phi}$ at step {\bf C} are mapped to
the complex moduli $U$ and $T$ corresponding to the complex structure
and the complex Kahler class of the $T^2$ along directions 1 and 5 at
step {\bf H}. The perturbative charge lattice $\Gamma_{(2,2)}$ in the
latter are given by the complex left and right moving momenta
\begin{eqnarray}
P_L &=& \frac{1}{\sqrt{2T_2 U_2}}[n_5 + m_5 TU + n_1 U - m_1 T]
\nonumber\\
P_R &=& \frac{1}{\sqrt{2T_2 U_2}}[n_5 + m_5 \bar{T}U + n_1 U - m_1 \bar{T}]
\label{plpr}
\end{eqnarray}
where $n_i$ are the KK momenta and $m_i$ are the windings along
directions 1 and 5. By the duality chain one also sees that for
$\chi=C_{(4)}=0$, 
$n_1$, $m_1$, $n_5$ and $m_5$ at step {\bf H} correspond to the
numbers of $D_1$, $D_{12345}$, $F_1$ and $NS_{12345}$ branes respectively.
Thus the real parts of $P_L$ and $P_R$ contain the information of the 
$F_1$ and $NS_{12345}$ charges while the imaginary parts  that of $D_1$
and $D_{12345}$ charges. Setting $n_5=m_5=0$ (i.e. setting the 
sources of $F_1$ and $NS_{12345}$ charges to zero), the real parts then
would carry the information of 
the respective induced charges in the presence of $\chi$
and $C_{(4)}$. For $C_{(4)}= \frac{p}{q} \chi$ with $p$ and $q$ some
integers, we see
that there are no induced charges of $F_1$ and $NS_{12345}$ branes
iff $n_1= \frac{p}{q}m_1$ \footnote{ For $\chi=1/2$ the equation 
$n_1= \frac{p}{q}m_1$ need be true modulo even numbers since we can
turn
on $n_5$ to set the real parts of $P_L$ and $P_R$ equal to
zero. Turning on $n_5$ would mean turning on a source of $F_1$ in the
original system. In particular this means that at the $\chi=1/2$ point
with
$p/q=Q_1$,
the bound state of 1 D5 and $Q_1$ D1 branes has a mass equal to the
sum
of the masses of
$(1 D5, (Q_1-2r) D1, r F1)$ bound state and  $(2r D_1,
-r F_1)$ bound state. Here the integer $r$ represents the source of $F1$
charge. It is not clear to us why the  existence of
such channels does not introduce singularity in the CFT}. 

In the usual type IIB on $T^4$ or $K3$, or the model $II$ considered
here, $\chi$ and $C_{(4)}$ are moduli fields and therefore one can
continuously go from the point $\chi=C_{(4)}=0$ to any other point. 
Furthermore,
since the subspace of the moduli space where the system is at
threshold is of real co-dimension greater than 1, 
one expects that the quantities
such as elliptic genus would not depend on the value of $\chi$. In the
models
$I$ and $III$, however, the $\chi$ and $C_{(4)}$ fields are projected out,
and therefore, the $\chi=1/2$ and/or $C_{(4)}=1/2$ point 
(which fortunately is invariant under
$\Omega$) cannot be connected to the trivial point. In fact, in
general the models obtained by $\Omega$ projection at $\chi=1/2$ may
be quite different from the ones at $\chi=0$. To illustrate this,
consider the usual type I' theory on $S^1 \times T^4$
(i.e. IIB$/\Omega I_4$ where $I_4$ acts on $T^4$ 
with no shift)
at $\chi=1/2$ and $C_{2345}=0$. 
By following the duality chain of table 1.2, at step
{\bf H}, this becomes IIB on $S^1 \times T^4/I_4$ where the effect
of $\chi$ is a $Z_2$ mixing of  $\Gamma_{(1,1)}$ and $\Gamma_{(4,4)}$ lattices
of momemta and windings on $S^1$ and $T^4$. More explicitly, $\chi
=1/2$  and $C_{(4)}=0$ implies that $Re(U)=1/2$ and $Re(T)=0$ in
eq(\ref{plpr}). 
The $I_4$ action which reflects the real parts of $P_L$ and $P_R$ is still
an automorphism of the resulting $\Gamma_{(5,5)}$ lattice. The set of $I_4$
invariant vectors now form a
sublattice $\Gamma$ of the self-dual lattice $\Gamma_{(1,1)}$ with
$\Gamma_{(1,1)}/\Gamma = Z_2$. By modular transformation from the
sector tr$I_4$ over the untwisted Hilbert space, one finds that in the
twisted sector the charge vectors are contained in the dual lattice 
$\Gamma^*$ of $\Gamma$ and the multiplicities are now given by
number of fixed points divided by the square root of the order of 
$\Gamma^*/\Gamma$ which in the present case gives $16/2=8$. In
particular, there will be 8 massless 5-dimensional gauge fields coming
from the twisted sector instead of the usual 16 (for $\chi=0$) when there is no
$Z_2$ mixing of $\Gamma_{(1,1)}$ and $\Gamma_{(4,4)}$. At the level of
type I' what this means is that out of the 16 orientifold fix planes, 12
come with positive charge while the remaining 4 come with negative
charge. As a result, there are only 8 pairs of space filling D5 branes
(instead of 16 in the absence of $\chi$). This is not surprising,
since by two T-dualities followed by an S-duality, type I' model is
mapped to IIB on $T^2$ modded by 
$\Omega (-1)^{F_L} I_2$ in the presence of a $Z_2$
discrete NS B field. The latter is known to give rise to 3 orientifold
planes with positive charge and 1 with negative charge, 
resulting in a rank
8 gauge group coming from the open string sector \cite{wittVS}. 
Note that the single D1 brane now is 
non-BPS, since it carries induced F1
charge. At step {\bf H} this is evident, since for
$(n_1,m_1)=
(1,0)$ the real parts of $P_L$ and $P_R$, which are reflected
by $I_4$ (and hence not associated to a conserved charge), are non-vanishing
and therefore the corresponding state must be non-BPS. These states
would decay to BPS states by emitting pairs of twisted states.
  
Let us now return to the D1/D5 system for model $III$, considered here at
$\chi=1/2$.
In the present case since $\Omega I_4$ is accompanied by a transverse shift
there are no additional gauge fields coming from the twisted sectors
in the U-dual theory at step {\bf H}. The multiplicities of the D1/D5
system can be read off from the orbifold group invariant 
untwisted states of the latter. Note that while $\chi =1/2$,
$C_{(4)}=0$
or $1/2$ (modulo integers), depending on whether $Q_1$ is even or
odd. At step {\bf H}, this means that while real part of the modulus
$U$ is $1/2$, the real part of $T$ is $0$ or $1/2$ in these two
cases. As a result, the corresponding charge vector is always in the
invariant sublattice $\Gamma$. Therefore the multiplicities of D1/D5
system
can be read off from the untwisted sector of the U-dual theory at step
{\bf H}.

For the 3 charge system, however, the U-duality map between the 
model $III$ at step
{\bf C} and model $II$ at step {\bf F} which exchanges D1 and KK charges 
will not give
much information. Indeed, assuming that the symmetric product CFT at
step {\bf C} describes the physics at $\chi=C_{(4)}/Q_1= 1/2$, the
U-duality
maps it to a system of 1 D5 brane and $Q_1$ units of KK charge in the
presence of $B^{NS}_{15} = C_{1234}/Q_1 =1/2$. The presence of such
background field would break the Lorentz invariance of the world sheet 
which is along 01 direction. Moreover, as we argued above, a single D5
brane is not BPS (for $C_{(4)}=Q_1/2$ with odd $Q_1$), since it has an induced 
$Q_1/2$ units of $F_1$ charge at step {\bf C}. At step {\bf
F} this implies that the single D5 brane carries 
$Q_1/2$ units of $F_5$ charge. Since this 
charge can be obtained by turning on 
an electric field along 05 directions in the D5 brane
world volume theory, it would
appear as the momentum mode of the Wilson line along the
5th direction. The induced charge therefore should mean a $Z_2$ shift in the 
momentum lattice of the Wilson line. Since the orbifold
group reflects this Wilson line, the shifted lattice for $Q_1$ odd
will have no invariant vector, implying that the single D5 brane is not
BPS. Moreover, since the D1
branes at step {\bf C} also carry the same induced charges in such a
way that, for $Q_1$ (modulo even numbers) D1 branes it cancels the 
induced charge of D5 brane, we conclude that there should be a shift
in the Wilson line lattice which is proportional to the total momentum
along direction 1. In a CFT the latter is given by $L_0-\bar{L}_{0}$.
Thus, there should be a very non-standard coupling between Wilson line 
lattice vectors and $L_0-\bar{L}_{0}$. These observations might give a
hint on how to go about modifying the symmetric product CFT to take
into account the $B^{NS}_{15} = C_{1234}/Q_1 =1/2$ background, 
although in the
following we will not attempt to do this.

\section{Effective World Volume CFT}

In this section we will try to obtain the effective fields theories for 
pure D5-branes, pure D1-branes and D1/D5 system for each of the models 
described in
the previous section that are related by U-duality.

\subsection{Transverse Shift}

We first discuss the case when the shift is transverse to the brane
system. In all the models we are considering we have $Z_2$ orbifolding of
type IIB theory compactified on $T^4 \times S^1 \times S^1$ where the $Z_2$ is
generated by an element of the form $g\cdot\sigma$ with $g$ being a
combination of $\Omega$ and reflection of $T^4$ and $\sigma$ is a shift on
the last $S^1$ factor. We are considering the system of D5-and D1-branes 
where the
D5-brane is wrapped on the $T^4$ and the first
$S^1$ factor and the D1 brane is wrapped on the first $S^1$ factor. If we 
have $Q_5$
D5-branes and $Q_1$ D1-branes in the quotient space, then in
the covering space there will be two identical sets of $(Q_5, Q_1)$ systems
which are placed at $X^6$ and $X^6 + \pi R_6$ with the world volume theory
on the two sets being identified via the $Z_2$ action $g$. Let $\Phi_1$
and $\Phi_2$ be the world volume fields on the two systems at
$X^6$ and $X^6 + \pi R_6$ respectively, then the identification is given
by $\Phi_2 = \hat{g} \Phi_1$, where $\hat{g}$ is the $Z_2$ action induced
by $g$ on the world volume fields. This essentially means that the
effective world volume theory is described by just one set of fields (say
$\Phi_1$) since the other set is not independent. This would just be the
field content of a single set of $(Q_5, Q_1)$ system in type IIB theory
compactified on $T^4 \times S^1 \times S^1$ with $R_6$ being the radius of
the last $S^1$. Where then does one see the effect of $Z_2$ orbifolding of
the underlying IIB theory ?  To understand this, note that among the fields
$\Phi_1$ there is one which corresponds to the 
common center of mass position of
D1/D5-branes along the $X^6$ direction. We shall denote this field by $X^6$.
As one changes the value of $X^6$, the entire system of the two sets of
branes moves along this direction. In particular when one moves $X^6$ all
the way to $X^6 +\pi R_6$ and the remaining fields $\Phi_1$ to $\hat{g}
\Phi_1$, then this system is equivalent to the original system (in a
description where one uses $\Phi_2$ as the independent field). Thus there
is a $Z_2$ gauging on the effective world volume theory defined by the
action $\hat{g}\sigma$, with $\sigma$ being the shift on the center of
mass
world volume field $X^6$. Now we will apply these general considerations to 
the cases of pure D5-branes, D1-branes (carrying KK-momenta) 
and D5/D1 system for the three models $I$, $II$ and $III$ respectively:
\vskip 0.1in
\noindent{\bf{D5-brane world volume theory}} \\

The low energy effective world volume theory on D5 branes 
in type IIB theory is
just the 6-dimensional ${\cal N}=(1,1)$ 
supersymmetric $U(Q_5)$ gauge theory. Let
$X^0, X^1,...,X^5$ be the directions along the D5-brane world volume, of which
the 4 directions $X^2,...,X^5$ are compactified on a torus. We can now carry 
out a
dimensional reduction so that the fields depend only on $X^0$ and $X^1$ 
\footnote{
We will throughout this
paper assume that the radii of this torus are of the order of string scale 
so that
in the low energy effective action one can ignore the KK modes for D5 branes 
as well
as the winding modes for the D1 branes along these directions.}. 
Let us denote by $\mu, \nu = 0,1$, $i,j =
2,...,5$ and $a,b = 6,...,9$. The world volume fields on the D5-branes are 
$A_{\mu}, A_{i}, X^{a}$ which are in the adjoint representation of $U(Q_5)$ and
their fermionic partners $\psi$. Sometimes for brevity of notation we will 
use $A_M$,
$M=0,1,...9$ to denote all the bosonic fields.  Let us denote by $g_1$, $g_2$ 
and
$g_3$ the $Z_2$ actions
$\Omega$, $I_{2345}$ and $\Omega.I_{2345}$ respectively. 
We start by determining the $\hat{g}_1$, $\hat{g}_2$, $\hat{g}_3$ induced
actions on worldvolume fields, once one accompanies the $Z_2$ elements
above with a geometric shift  $X^6 \rightarrow X^6 + \pi R_6$.
For the $(4,4)$ model this follows directly from the interpretation
of $A_i$ as describing the Wilson lines along $T^4$
and from supersymmetry:  
\be
\hat{g}_2 :~~~~~ X^6 \rightarrow X^6 + \pi R_6 , ~~~ A_i \rightarrow - A_i,~~~
\psi \rightarrow \Gamma_{2345} \psi 
\label{g2}
\ee
To understand the action $\hat{g}_1$, let us reconsider 
the system of D5-branes.
We have a total of $2 Q_5$ D5-branes 
in the covering space, where $Q_5$ of them are
sitting at the center of mass position $X^6$ and the remaining $Q_5$ at $X^6 +
\pi R_6$. Thus in the resulting system $U(2Q_5)$ is broken to 
$U(Q_5)\times U(Q_5)$.
The gauge fields can then be represented in terms of $Q_5 \times Q_5$ blocks:
\be
{\cal A}_M = \left(\begin{array}{cc} A_M & 0 \\ 0 & A'_M \end{array}\right),
~~~~~ \Psi = \left( \begin{array}{cc} \psi & 0 \\ 0 & \psi' \end{array}
\right)
\ee
where $A_M$ and $A'_M$ are the $U(Q_5)$ 
gauge fields on the two sets of branes. These two
gauge fields are of course not independent 
of each other; they should be related by
the $Z_2$ action in the underlying string theory. The shift exchanges the
two sets of branes and therefore exchanges $A_M$ 
with $A'_M$, while the $\Omega$
symplectic projection acts on the Chan-Paton indices with the result 
\be
{\cal A}_M = \mp \Omega_5 {\cal A}_M^t \Omega_5, ~~~~ \Psi = -\Gamma^{(7)}\Omega_5 \Psi^t 
\Omega_5   
\ee
with $\Omega_5$ written in terms of $Q_5 \times Q_5$ blocks as
\be
\Omega_5 =i\left( \begin{array}{cc} 0 &{\bf 1} \\  
-{\bf 1} & 0 \end{array} \right)
\ee 
and $\mp$ sign means $-$ for $M= 0,...,5$ and $+$ for $M=6,...,9$. 
In the above
$\Gamma^{(7)}= \Gamma_{012345}$. More
explicitly the above projection implies
\be
A_{\mu}' = -A_{\mu}^t \quad  A_{i}' = -A_{i}^t \quad X'_a = X^t_a \quad   
\Psi ' =- \Gamma^{(7)} \Psi^t  
\ee
We can now take the independent set of fields to be $A_M$ and $\psi$. 
However the
$Z_2$ action which takes the values of these fields to that of $A'_M$ and 
$\psi'$
will give rise to a configuration which is indistinguishable from the 
original one.
Thus the induced $Z_2$ gauging on the set of fields $A_M$ and $\psi$ is 
given by

\be
\hat{g}_1 : A_{\mu} \rightarrow - A_{\mu}^t,  ~~~ A_i \rightarrow -A_i^t, 
~~~ X_a
\rightarrow
X_a^t  , ~~~
\psi \rightarrow -\Gamma^{(7)} \psi^t
\label{g1}
\ee

Finally since $g_3 = g_1\cdot g_2$ it follows that the induced action 
$\hat{g}_3$ 
is
given by:
\be
\hat{g}_3 : A_{\mu} \rightarrow - A_{\mu}^t,  ~~~ A_i \rightarrow  +A_i^t, 
~~~ X_a
\rightarrow
X_a^t , ~~~
\psi \rightarrow -\Gamma_{01} \psi^t
\label{g3}
\ee

In the Coulomb branch the moduli space 
is given by taking diagonal matrices for 
$A_i, X_a$ and $\psi$, up to the Weyl group which is the permutation group 
$S_{Q_5}$ acting on the $Q_5$ eigenvalues. In particular, the diagonal 
entries in 
$X_a$ describe the transverse positions of each of the D5-branes. 
If one moves any
one of these branes along the $X_6$ direction by an amount $\pi R_6$ and 
simultaneously 
changes the field on that brane by the action of $\hat{g}$, then this system 
would be
indistinguishable from the original one. 
Thus the conformal field theory
describing the Coulomb branch can be written as
\be
{\cal M}_{\rm coulomb}=
(R^3 \times S^1 \times T^4)^{N}/S_{N}\ltimes Z_2^{N} 
\label{mc}
\ee
where $N=Q_5$, $S^1$ denotes the circle 
along $X^6$ direction and $T^4$ (dual of the
original $T^4$), appearing from the Wilson lines,  
will be coordinatized by $A^2,...,A^5$. 
The $Z_2^{Q_5}$ orbifold group is generated in models
$I$, $II$ and $III$ by  
(\ref{g1}), (\ref{g2}) and (\ref{g3}) respectively, acting on each 
of the $Q_5$ copies of $R^3 \times S^1 \times T^4$. Of course 
this  $Z_2^{Q_5}$ action does not commute with the
permutation group $S_{Q_5}$ and
indeed in (\ref{mc}) we have a  semidirect product.
This may seem a bit puzzling since $S_{Q_5}$ is
the remnant of the $U(Q_5)$ gauge symmetry of the system before going to the
Coulomb branch. The point is that at the level of $U(Q_5)$ gauge theory, 
where one 
ignores the massive modes coming from the strings stretched between 
the D5-branes at $X^6$ and their images 
at $X_6 +\pi R_6$, this $Z_2^{Q_5}$ 
symmetry is broken. 
However it is easy to see that this should 
be the symmetry of the theory
when one  includes these massive states that transform
in the $(Q_5, \bar{Q}_5)$ representations of $U(Q_5)\times U(Q_5)$ gauge 
symmetry. 
In the infrared limit in the Coulomb branch, when one ignores all the massive
off-diagonal modes, the theory has manifest $Z_2^{Q_5}$ symmetry.

Acting on the transverse fields entering in (\ref{mc}), the 
induced $Z_2$ actions (\ref{g1}),(\ref{g2}) and (\ref{g3}) can be
written as
\bea
\hat{g}_1 &=&(-)^{F_L}\, I_{2345} \, \sigma_{p_6}\nonumber\\
\hat{g}_2 &=& I_{2345} \, \sigma_{p_6}\nonumber\\
\hat{g}_3 &=&(-)^{F_L}\, \sigma_{p_6}
\label{gs}
\eea
where we recognize in $\Gamma_{01}$ the world sheet chirality 
operator $(-)^{F_L}$ and in $I_4$ and $\Gamma_{2345}$ the reflection
operator acting on $T^4$'s bosons and fermions respectively.  

Note that for $Q_5=1$ the effective theories we have obtained are just the 
type IIB $Z_2$ orbifolds generated by 
(\ref{gs}) in the static gauge. Indeed from the
U-duality map table 1.1, we recognize the corresponding 
fundamental sides of the D5-KK systems
in the string vacua defined by the $\Omega$, $I_4$ and $\Omega I_4$
projections respectively. 

Note that $Z_2$ orbifolding breaks the 
(8,8) supersymmetry of the parent IIB 
system down to (4,4) for models $I$ and $II$, while 
it breaks to (8,0) for model $III$. This is to
be expected since by T-dualities along 2,3,4 and 
5 directions the D5-brane in the 
model $III$ is mapped to D1-brane in model 
$I$, which is just a type I-like theory.

\vskip 0.1in
\noindent{\bf{D1-brane world volume theory}} \\

One can follow the above reasonings also for D1-branes. The only difference 
is that the 
$\Omega$ projection now gives an extra negative sign also for $2,3,4$ and $5$
directions since, unlike the D5 brane case, they are now transverse.
It is easy to see that this just exchanges $\hat{g}_1$ with $\hat{g}_3$ while
leaving the $\hat{g}_2$ unchanged. Once again for $Q_1=1$ this reproduces the
fundamental side (under column D), for these three cases. 
Moreover, this is also
consistent with T-duality since by four T-dualities along 2,3,4 and 5 
directions, 
the D5 brane is exchanged with D1 brane and model $I$ is exchanged with 
model $III$.

\begin{itemize}

\item{We summarize the various $Z_2$ actions entering in (\ref{mc}) 
for pure D5- ($N=Q_5$) and pure D1- ($N=Q_1$) brane systems
in the following table:\\\\
\begin{tabular}{ccccc}
Model &  &   $D1-KK$ & &      $D5-KK$ \\
$I$     &  &   $ (-)^{F_L}$ & & $(-)^{F_L} \cdot I_4$  \\
$II$    &  &   $  I_4 $ & & $ I_4$ \\
$III$   &  &   $(-)^{F_L} \cdot I_4 $ & & $(-)^{F_L}$ \\
\end{tabular}\\\\
{\it Table 2.1: D1-KK, D5-KK bound states: Induced $Z_2$ actions entering
in (\ref{mc})}  \\}

\end{itemize}

In the comparison with the fundamental string multiplicities,
we will restrict ourselves to two-charge systems and 
therefore we should focus on the untwisted sectors of
(\ref{mc}). 
$Z_2$-twisted states correspond, in each case, to
states carrying additional charges corresponding to
D1(D5)-branes oriented along
the direction $X_6$. In the dual picture, 
this corresponds to a fundamental string
carrying winding-momentum along $X^1$, together with
an additional odd winding charge $F_6$. 
Indeed degeneracies for fundamental string states carrying
odd $F_6$ winding charges can be read off from the 
twisted sector amplitudes in {\bf A} or {\bf D} and with
a little effort can be seen to match the ones coming
from the three-charge bound states. 
In the following we will however limit ourselves to 
the two-charge system, and (\ref{mc}) will be always 
understood (for transverse shifts) restricted to the  
$Z_2$-untwisted sector.

\vskip 0.1in
\noindent{\bf{D1/D5 system}}

Now let us add $Q_1$ D-strings to the system of $Q_5$ D5-branes.
D1-branes are along the $X^1$ direction while D5-branes are along 
the $X^1,...,X^5$ directions. In the covering space again this
system would split into two copies of the D1/D5 system sitting at $X^6$ and
$X^6+ \pi R_6$. For each of these two sets, the 1+1 dimensional common
world volume theory is just a supersymmetric sigma model on the moduli
space $\cal{M}$ of $Q_1$ instantons of 
${\cal N}=4$ $U(Q_5)$ gauge theory on $T^4$,
times the center of mass fields corresponding to the common transverse
directions $R^3 \times S^1$. This is exactly the model appearing in the
type IIB context. 
Since each of the $Z_2$ actions on the D5-brane world volume gauge
fields described above leaves the self-duality equations invariant, it 
follows that it induces an
action $\hat{g}$ on the instanton moduli space $\cal{M}$.  Following the logic
described above,
it then follows that the effective world volume theory is the $Z_2$
gauging of the theory in the IIB case where the $Z_2$ is generated by
$\hat{g}\cdot \sigma$.

In the infrared limit the (4,4) supersymmetric sigma model 
on ${\cal M} \times R^4$ 
would flow to a (4,4) CFT. It is conjectured that this CFT
is a symmetric product space $R^4 \times T^4 \times (T^4)^N/S_N$. There has 
been
much critical discussion of this conjecture in the literature 
\cite{W1,SW1,LM}. 
As mentioned in the previous section, it is
generally believed that this is true for the 
$Q_5=1$ case at the point $\chi = C_{2345}/Q_1 = 1/2$ and
$V_4=Q_1$ in
the moduli space, where $\chi$ and $C_{2345}$ are the RR 0-form and
4-form
fields and $V_4$ is the volume of the $T^4$ along 2345 directions
\footnote {$V_4$ is set at $Q_1/Q_5$ in order to minimize the mass of the bound
state.
This amounts to setting the asymptotic value of the $V_4$ equal to its
fixed value at the horizon of the D1/D5 system. Note that precisely
for this choice the supergravity solution admits a constant $\chi$ 
and $C_{2345}$ \cite{dmwy}.}
For other  values of $Q_5$ with $Q_1$ relatively coprime, 
this CFT with $N=Q_1\cdot Q_5$ 
perhaps describes the system at some point in the moduli space  of the
IIB theory on $T^4$, which
is related to the point for $Q_5=1$ case by a U-duality that maps the 
$(Q_5, Q_1)$ system 
to the $(1,Q_1\, Q_5)$ system \cite{LM}. Note that the point 
$\chi= C_{2345}/Q_1=1/2$, at which the theory is described by the
symmetric product CFT, is invariant under the action of $\Omega$
and $I_4$. In model $II$, $\chi$ and $C_{2345}$ are moduli fields and
therefore
one can continuously go from this point to the point where these
fields are switched off, and one expects that the quantities
such as elliptic genus will not change in this process. However, for
models $I$ and $III$, these moduli fields are projected out and as a
result these models are frozen at the $\chi=C_{2345}/V_4=1/2$
point. Therefore,
the elliptic genus computed at this point may not be the same as the
one at the trivial point. We will comment on this in the conclusions,
but
for the rest of the paper we will be working at the point in the
moduli space where the symmetric product CFT is the correct description.

For $Q_5=1$ the
various factors appearing in the CFT have a clear interpretation: the 
D-flatness
condition sets all the bifundamental fields coming from 1-5 open string 
states to
zero, leaving behind only the Cartan directions of the 1-1 $U(Q_1)$ adjoint 
states.
The latter  have the interpretation of the positions of the D-strings inside 
$T^4$.
Thus the factor $(T^4)^{Q_1}/S_{Q_1}$ represents the positions of the $Q_1$
instantons, while the center of mass factors $R^4$ and $T^4$ represent the
transverse position of the D5 brane and its $U(1)$ Wilson lines on the 
$T^4$ (so
more precisely this should be the dual torus). In our case, of course since the
transverse direction along $X^6$ is compactified on a circle, $R^4$ should be
replaced by $R^3 \times S^1$.

With the physical interpretation of the various fields appearing in the CFT 
being
clear, we are now in a position to deduce the induced $Z_2$ action on the
instanton moduli space for each of the
three models. Let us denote by $A_i$ for $i=2,...,5$ the four $U(1)$ Wilson 
lines
of the D5-brane gauge field
and by $X_i^{(\ell)}$ for $\ell = 1,...,Q_1$ the positions of the $Q_1$ 
instantons
on $T^4$. Finally we denote by $X_a$ for $a=6,...,9$ the coordinates of 
the center of mass
transverse position $S^1 \times R^3$. The little group 
$SO(4) \equiv SU(2)_A \times SU(2)_Y$ acts on the tangent space of 
$S^1\times R^3$. In the type IIB theory the
resulting CFT has (4,4) supersymmetry. The left and right moving supercharges 
come with definite chiralities with respect to the little group $SO(4)$. 
Specifically  the left moving supercharges are two $SU(2)_A$ doublets while the
right moving ones are two $SU(2)_Y$ doublets. The supermultiplets then are

\begin{tabular}{ccc}
Bosons  &  Left-moving Fermions & Right-moving Fermions \\
$X_a \equiv X_{AY}$ & $\psi_A$ & $\tilde{\psi}_Y$ \\
$A_i$ &$\psi_Y$ & $\tilde{\psi}_A $\\
$X_i^{(\ell)}$ & $\psi_Y^{(\ell)}$ & $\tilde{\psi}_A^{(\ell)}$ \\
\end{tabular}

$\hat{g}_1$ leaves $X_a$ and $X_i^{(\ell)}$
invariant, since these are respectively the center of mass position 
and the positions
of D1-branes in $T^4$. It however takes $A_i$ to $-A_i$, since $\Omega$ 
projects out
the $U(1)$ gauge field. The easiest way to understand its action on 
the fermions is
to use the fact that in this theory the D1/D5 system should preserve (4,0)
supersymmetry. As a result, $\psi_A$ and $\psi_Y^{(\ell)}$ should remain 
unchanged
while $\psi_Y$ must pick up a minus sign. On the right-moving fermions the 
action is
exactly the reverse of this, i.e. $\tilde{\psi}_Y$ and 
$\tilde{\psi}_A^{(\ell)}$ 
should pick up a minus sign while $\tilde{\psi}_A$ 
should remain unchanged. This 
is
because $X^a$ and $A_i$ are D5-brane fields and as one can see
from table 2.1 or eqs. (\ref{gs})   
$\hat{g}_1$  acts on
the fermions 
by $\Gamma^{(7)}=(-)^{F_L} I_4$.
Thus $SU(2)_A$ and $SU(2)_Y$ doublets must appear with opposite 
signs.
On the other hand, $X^{(\ell)}_i$ are the D1-brane  fields and on the fermions
$\hat{g}_1$ acts as $\Gamma_{01}$. Thus the left and right moving fermions 
appear with opposite signs. To summarize $\hat{g}_1$ maps
\begin{eqnarray}
(X_a, X_i^{(\ell)}, \psi_A, \psi_Y^{(\ell)}, \tilde{\psi}_A)
&\rightarrow&
(X_a + \delta_{a6}\pi R_6, X_i^{(\ell)}, \psi_A, \psi_Y^{(\ell)}, 
\tilde{\psi}_A)
\nonumber\\
(A_i, \psi_Y, \tilde{\psi}_A^{(\ell)}, \tilde{\psi}_Y)
&\rightarrow& -(A_i, \psi_Y, \tilde{\psi}_A^{(\ell)}, \tilde{\psi}_Y)
\end{eqnarray}

In model $II$ the induced action is more straightforward to see. In this case
the D1/D5 system preserves the full (4,4) supersymmetry of the 
parent type IIB system.
Thus it is sufficient to specify the $\hat{g}_2$ action on the bosonic fields.
Since $g_2$ is the inversion $I_{2345}$, it follows that it gives negative 
sign to
$A_i$ and $X_i^{(\ell)}$ and all their fermionic partners.
Finally, $\hat{g}_3$ is just obtained as product of $\hat{g}_1.\hat{g}_2$. 
We conclude then that D1/D5 gauge theories flow in the infrared
to orbifolds CFT of the type 
\be
{\cal M}_{\rm higgs}=\left( R^3\times S^1\times T^4
\times (T^4)^N/S_N \right)/Z_2
\label{mh}
\ee
with $N=Q_1 Q_5=Q_1$ and $Z_2$ acting diagonally 
in the way specified in the
following table:

\begin{tabular}{ccc}
Model & &  D1-D5 \\
$I$     & &$ (-)^{F_L}\, I_4^{\rm c.m.}$ \\
$II$    & &$  I_4^{\rm c.m.}\,I_4^{\rm sp} $ \\
$III$   & &$(-)^{F_L}\, I_4^{\rm sp}$ \\
\end{tabular}\\\\
{\it Table 2.2: D1/D5 systems: $Z_2$ actions entering in (\ref{mh})}  \\

We denote by $(-)^{F_L}$ the total left moving 
fermionic number, $I_4^{\rm c.m.}$
the reflection of the fields corresponding to the first $T^4$ 
factor in (\ref{mh}) and 
by $I_4^{\rm sp}$ the diagonal $Z_2$ reflecting bosonic and fermionic
fields in the symmetric product.

Let us finally observe that, both 
in models $I$ and $III$ the $T^4$ components
of the NS-NS B-field,
as well as the $T^4$ components of the 
R-R 4-form and the axion, are projected
out from the massless spectrum, due to the $\Omega$-projection.
But in the $T^4$ case the self-dual part of the above
B-field and a combination of four-form and axion are
the moduli that, when switched on, render
the D1/D5 system a bound state below threshold. 
They in turn correspond,
in the effective gauge theory to Fayet-Iliopoulos D-terms
and theta-term for the $U(1)$ gauge field. 
The effective gauge theories
derived in  sub-section 3.1 agree with the spectrum
above, in the sense that  in models
$I$ and $III$, due to the
$\Omega$ action on Chan-Paton factors, 
there is no room for Fayet-Iliopoulos 
D-terms in the potential for hypermultiplets
and/or theta-term for the $U(1)$ gauge field.
(the $U(1)$ generators are traceless).
On the other hand, in model $II$ the D1/D5
gauge theory agrees with the fact that the above 
moduli are actually there. 

\subsection{Longitudinal Shift}

We now consider the shift along the common world volume direction $X_1$.
The general discussion in this case is very similar to the transverse
shift case. If $\Phi$ denotes the set of world volume fields in the type IIB
theory, and $\hat{g}$ the induced $Z_2$ actions, then the 
worldvolume fields satisfy the condition:
\be
\Phi(X_1 + \pi R_1 , X_0) = \hat{g} \Phi(X_1, X_0)
\ee
Taking the interval of $X_1$ to be $\pi R_1$, what this condition says is
that the fields on which $\hat{g}$ acts as $-1$ are anti-periodic along 
the $X_1$ direction and the ones on which the $\hat{g}$ action is +1 are
periodic. This means that the CFTs are again given by 
(\ref{mc}), (\ref{mh}) with $Z_2$ actions specified by tables 2.1 and
2.2, but with the twists now oriented in the $\sigma$-direction,
i.e. the $Z_2$ twisted sectors of (\ref{mc}), (\ref{mh}).
Notice that, in contrast with the transverse case, this represents
still a two-charge system.  

Another way to see that the above proposal must be correct, is to start from
the effective field theories in the transverse case and consider the
threshold corrections due to the single (i.e. minimal unit)
D5- or D1-instanton, obtained by wrapping the time
directions of these systems
on the $X_6$ circle by a length $\pi R_6$ (i.e. half winding). The resulting
amplitude is just the
one loop amplitude in the orbifold sector given by the 
insertion of the operator
$\hat{g}$. This is
because in the path-integral formulation, it is in 
this sector that there is a half
winding, corresponding to the shift along the 
time direction. On the other hand,
by a modular transformation, we can exchange $t$ and $\sigma$, 
and the resulting 
path integral should be interpreted as that of the field theory living in the
single D5-aligned along directions 23456 or a 
single D1-brane along direction 6, with
longitudinal shift along $X_6$. This field theory is just the twisted
sector ($\hat{g}$ twist along $\sigma$ direction) of the $Z_2$ orbifold of $R^4
\times T^4$. Putting $N$ copies of these together 
should, in the Coulomb branch,
reproduce the conjecture of the previous paragraph. 
In the case of the D1/D5 system the
same argument applies in a more straightforward 
way since there is only a single
copy of the center of mass $R^4$. We conclude then that in the
longitudinal shift case the CFT descriptions are given again in
terms of (\ref{mc}), (\ref{mh}) but now are the $Z_2$-twisted sectors 
which are relevant to our discussion.  

There is however an apparent puzzle we would like to discuss here.
Consider D1-brane in model $I$. $\Omega\sigma_p$ projects the $U(N)$ group
to $SO(N)$ so that $SO(N)$  gauge fields are periodic while the remaining
ones, that are in
the symmetric  tensor representation of $SO(N)$, 
are anti-periodic. Now let
us put D5-branes. The Gimon-Polchinski consistency condition \cite{gp}
would at first sight imply that the $\Omega$-projection on D5-brane
Chan-Paton factors
should be chosen to be symplectic. This would mean
a doubling phenomenon, i.e. one would need an even 
number of type IIB D5-branes. 
If this would be the case, then the whole towers of states
with odd windings at step {\bf H} would be missing in the dual
description.

There are various ways of seeing 
the Gimon-Polchinski consistency condition. One of 
them involves a consideration of
the Dirac charge quantization condition. This was one
way to see that, in the usual type I theory, 
if there is a  single D-string,
then D5-branes should be paired (with respect to type 
IIB counting) because of
the fact that in the 1-5 sector there is a factor of 1/2 due to the
$\Omega$
projection. 
Indeed, this additional factor rescales the D1, D5 charges by 
$1/\sqrt{2}$ with respect to their type IIB cousins, and the Dirac
quantization condition $Q_e Q_m=2\pi$ requires therefore the 
claimed pairing of D5-branes.  
Now let us discuss our case. Consider the D5-brane to be
longitudinal to the direction $X_1$ along which the shift acts. Then the
Poincare` dual B-field, which enters in Dirac 
quantization condition, would
refer to the D-string which is transverse to $X_1$. 
But in this case we have
actually 2 D-strings (one sitting at say $X_1$ and the other at $X_1 +
\pi R_1$). So quantization condition is satisfied just with one 
D5-brane wrapped on the circle 
with circumference $2\pi R_1$. Similarly, if 
one takes the D-string longitudinal to $X_1$, 
then its Poincare` dual involves a
D5-brane that is transverse to $X_1$, in which case again we have 2 
D5-branes, showing that it suffices to have just one D-string.

The other way to see the appearance of this condition
is to consider the action of $\Omega^2$ on the open string
states. In the usual Type I theory, Gimon-Polchinski showed \cite{gp}
that on the 
1-5 open strings $\Omega^2$ picks up an extra minus sign, due to the fact
that these states involve a twist field along the four directions
longitudinal to D5-brane and transverse to the D1-brane. Including the
action of $\Omega$ on the Chan-Paton indices we have
\be
\Omega^2:  |\alpha, \mu> \rightarrow  -(\gamma^t \gamma^{-1})_{\alpha
\beta} | \beta, \nu> (\gamma'^t \gamma'^{-1})_{\nu \mu}
\ee
where $\alpha, \beta$ and $\mu, \nu$ are the Chan-Paton indices on D1- and
D5-branes respectively and $\gamma$ and $\gamma'$ are $\Omega$ actions on
the D1- and D5-brane Chan-Paton indices. Due to the extra minus sign above,
one concludes that if $\gamma$ is symmetric then $\gamma'$ must be
anti-symmetric and vice versa. So, if one system is projected onto
the Orthogonal Group then the other must be projected onto the 
Symplectic Group.  

In our case however $\Omega$ is accompanied by the shift $\sigma_1$. Thus
$g_1^2 = \Omega^2 \sigma_1^2$ and we can take both systems to have
Orthogonal projections provided $\sigma_1^2 =-1$ on the 1-5 string states.  
This means that 1-5 string states will carry half-integer momenta along
the $X_1$ circle. Since 1-5 states are bi-fundamentals this can be 
thought of as  turning on of a $Z_2$ Wilson line in one of the systems
along the circle. 

In Appendix A we construct the open string theory living on 
D1/D5 system in the presence of a longitudinal shift.
We follow the open string descendant techniques developed in 
\cite{torvergata}.
The consistency condition translates in this formalism in
the requirement that Klein bottle, Annulus and Moebius
amplitudes in the transverse channel admit a sensible interpretation
as closed string exchanges between boundaries (D-branes) 
and crosscaps (orientifold planes).
Once again, one finds that the Orthogonal assignements for both D1 and D5
Chan-Paton indices are allowed provided that 1-5 string states 
carry half-integer momenta along the $X_1$ circle.
This is the same condition that we found in the previous paragraph. 

This result might seem a little surprising, since the D1/D5 system can be 
thought of as Y-M instantons in the D5-brane world volume. It is known 
that in the context of the ADHM construction, $SO(N)$ instantons have $Sp(k)$ 
symmetry where $k$ is the instanton number, and vice versa. This is indeed 
the result for the standard $\Omega$ projection.
In our case however $\Omega$ is accompanied by a shift. What this means 
is that we are looking for $U(N)$ instantons in the 4 directions 
spanned by $X_2,....X_5$, and the moduli of the instantons are slowly 
varying functions of $X_1$ in such a way that
\be
A_{\mu}(X_1 + \pi R_1) = - g^{-1}A_{\mu}^t(X_1)g  
~~~~~~~~\mu=2,3,4,5
\label{aproj}
\ee
where $g \in U(N)$ is a slowly varying function of $X_1$. 
The periodicity condition 
(up to a possible Wilson line $h \in U(N)$) as 
$X_1 \rightarrow X_1 + 2\pi R_1$
implies 
\be
g^*\cdot g = h
\ee

For orthogonal and symplectic projections $h=+ {\bf{1}}$
and $h=-{\bf{1}}$ respectively and in these two cases we can take
$g$ to be $+{\bf{1}}$ and the symplectic matrix ${\bf J}$ respectively.
In the latter case of course $N$ must be even.

Let us now see how this condition is translated on the ADHM data 
(here we will take
the 4-dimensional space where the instanton is sitting to be $R^4$ since 
the discussion of the doubling phenomenon should not depend on whether 
the space is $T^4$ or $R^4$). The ADHM data consists of a 
$(N+2k) \times 2k$ matrix $\Delta$ defined as
\be
\Delta_{\lambda, i \dot{\alpha}} = a_{\lambda, i \dot{\alpha}} + 
b_{\lambda, i}^{\alpha} x_{\alpha \dot{\alpha}}
\ee
where $x_{\alpha \dot{\alpha}} = x_{\mu} \sigma^{\mu}_{\alpha \dot
{\alpha}}$ and the 
indices $\lambda = u + j \beta$ with $u$ running over $N$ indices and 
$i,j$ run over $k$ indices. $\Delta$ satisfies the quadratic constraint
\be
\bar{\Delta}^{\dot{\alpha}}_{i,\lambda} \Delta_{\lambda, j \dot{\beta}}
=\delta^{\dot{\alpha}}_{\dot{\beta}} f^{-1}_{ij}
\label{quad}
\ee
where $f$ is a $k\times k$ hermitian matrix.

The self-dual gauge fields are then given by
\be
A_{\mu} = \bar{U} \partial_{\mu}U
\ee
where $U$ is an $(N+2k) \times N$ matrix which satisfies the equations
\be
\bar{U}U={\bf{1}}, ~~~~~~~~~~ \bar{\Delta}U = \bar{U}\Delta=0
\label{ADHMcon}
\ee
From the above it follows that the projection operator 
$U\bar{U} = {\bf{1}}-\Delta f \bar{\Delta}$. 

The symmetries of these equations are 
\be
U \rightarrow B\cdot U\cdot g, ~~~~~~~~~~~~~~\Delta
\rightarrow B\cdot \Delta\cdot (C \times {\bf{1}}_{2\times 2}) ,
\label{adhmsym}
\ee
where $g$ is a local
$U(N)$ transformation while $B$ and $C$ are 
independent of $X_{\mu}$ and are in
$U(N+2k)$ and $GL(k)$ respectively. 
Using this freedom in defining the data, we can
set $b_{u,i}^{\alpha}=0$ and $b_{j \beta, i}^{\alpha}= \delta_{ij}
\delta_{\beta}^{\alpha}$. Then the instanton moduli are contained in the 
matrix $a$ which, as follows from eq.(\ref{quad}),
satisfies the constraint that $a^{\mu}_{ji}$ is in the adjoint 
representation of $U(k)$,
where $a^{\mu}$ is defined via $a_{j \beta, i \dot{\alpha}} =
a^{\mu}_{ji}\sigma^{\mu}_{\beta \dot{\alpha}}$. There are also the 
$3k^2$ D-term constraints that are quadratic in $a$, that follows from
(\ref{quad}), but they will not 
concern us here. With this canonical choice 
for $\Delta$, the global symmetry group 
$U(N + 2k) \times GL(k)$ reduces to $U(N)\times U(k)$. Explicitly,
this corresponds to taking $C$ in eq. (\ref{adhmsym}) to be in $U(k)$ and
\be
B = \left(\begin{array}{c} D \\ C^{-1}\times {\bf{1}}_{2\times2}  
\end{array}\right)
\nonumber
\ee 
with $D \in U(N)$. 
Note that the moduli $a^{\mu}_{ij}$ transforming in the 
adjoint representation of $U(k)$ are part of the 1-1 string states
that define the position of the D1-brane inside the D5-brane, while 
$a_{u, i \dot{\alpha}} \equiv w_{u, i \dot{\alpha}}$ are 
the 1-5 string states that are bi-fundamentals of
$U(N)$ and $U(k)$. The spinorial index $\dot{\alpha}$ just refers to
the fact that the bosonic 1-5 string states are spinors of the
$SO(4)$ acting on $X^2,...X^5$. 

However at this point we still have two $U(N)$ actions:
the local $U(N)$ action on $U$ on the right and the global $U(N)$ action
on the left. The instanton gauge field $A_{\mu}$, which lives on the 
D5-brane sees only the local action, while the ADHM
data $w$ see only the global action. 
In order to relate this system to the D1/D5
system, we must identify these two $U(N)$ actions.  
The basic 
point is to choose a particular gauge for the instanton 
solution that fixes the 
local $U(N)$ symmetry. We will choose the 
singular gauge \cite{dorey} which is described as follows.
Writing 
$U$ as an $N\times N$ block $V$ and $2k \times N$ block $U'$, the condition
$U\bar{U} = {\bf{1}}-\Delta f \bar{\Delta}$ 
implies $V \bar{V} = {\bf{1}}-w f \bar{w}$. 
Given a solution for $V$, $V\cdot g$ will also solve this 
equation, for $g$ being a local $U(N)$ transformation. 
Choosing the singular gauge amounts to taking
$V$ to be one of the $2^N$ matrix square roots of the right-hand side 
$({\bf{1}}-w f \bar{w})^{1/2}$. With this choice, it is clear that a 
transformation
$w \rightarrow D w$ implies $V \rightarrow D V D^{-1}$ and the two $U(N)$'s are
identified. 

Let us now return to the $Z_2$ projection condition (\ref{aproj}). With the two
$U(N)$ actions identified, this condition on the ADHM data becomes:
\begin{eqnarray}
w(X_1 +\pi R_1) &=& g w^*(X_1) (C \times \sigma_2)  \nonumber \\
a_{\mu}(X_1+\pi R_1) &=& C^{-1} a_{\mu}^*(X_1) C
\end{eqnarray}
where $C \in U(k)$. Here we have used the fact that 
$x = \sigma_2 x^* \sigma_2$.
Repeating this equation twice we find:
\bea
w(X_1 + 2\pi R_1) = - (g\cdot g^*) 
w(X_1) (C^*C), ~~~~a_{\mu}(X_1+2\pi R_1)
=   (C^* C)^{-1} a_{\mu}(X_1) (C^* C) .
\nonumber
\eea
As stated earlier, orthogonal and symplectic 
projections of $U(N)$ correspond to
$g={\bf 1}$ and $g={\bf J}$ respectively. Similarly, the
orthogonal and symplectic projections of 
$U(k)$ are given by choosing $C={\bf 1}$ and $C={\bf J}$ respectively. 
The above equations show
that if both the groups are projected to Orthogonal 
or Symplectic Groups, then 
$w$, which represents 1-5 string states, are anti-periodic as $X_1\rightarrow
X_1+2\pi R_1$, while if they are projected in the opposite ways the $w$ are
periodic.
This is exactly the condition
we found using Gimon-Polchinski consistency condition. 

It is instructive to consider $k=1$, since in this case we can 
explicitely solve the ADHM constraints. The result is\cite{dorey}:
\be
w_{u \dot{\alpha}} = \rho G\left(\begin{array}{c}  {\bf{0}}_{[N-2]\times[2]} \\
{\bf{1}}_{[2]\times[2]}
\end{array}\right), ~~~~~~ G \in \frac{SU(N)}{SU(N-2)}
\label{aualpha}
\ee
where $\rho$ is the scale of the instanton.
One can then solve for $U$ in the singular gauge satisfying equation 
(\ref{ADHMcon}) and obtain the gauge
field as:
\be 
A_{\mu} = G \left(\begin{array}{cc} 0 & 0 \\ 0 & A_{\mu}^{SU(2)}  
\end{array}\right) G^{-1}
\ee
where $A_{\mu}^{SU(2)}$ is the standard $SU(2)$ single instanton gauge field
in the singular gauge with scale $\rho$ and position $a^{\mu}$:
\be
A_{\mu}^{SU(2)} = \frac{\rho^2 \bar{\eta}^c_{\mu \nu} (X-a)^{\nu} \sigma^c}
{(X-a)^2 ((X-a)^2 +\rho^2)} 
\ee

The moduli of the instantons are the position $a_{\mu}$, 
the scale $\rho$ and the 
gauge orientations contained in $G$. These moduli are 
now slowly varying functions
of $X_1$ in such a way that the $Z_2$ projection condition (\ref{aproj}) is
satisfied.
It is easy to see that this condition implies that
\begin{eqnarray}
a^{\mu}(X_1+ \pi R_1) &=& a^{\mu}(X_1), ~~~~~~~\rho(X_1+ \pi R_1)=\rho(X_1),
\nonumber \\
G(X_1+ \pi R_1) &=& g G^*(X_1) \sigma_2
\end{eqnarray}
Repeating this twice and using  eq. (\ref{aualpha}), we find the condition 
$w(X_1+ 2\pi R_1) = -(g g^*) w(X_1)$.
Taking the orthogonal projection for $U(N)$, namely $h={\bf 1}$, 
we recognize from above
that the ADHM data $w$, which represents the 1-5 string states
are anti-periodic as $X_1 \rightarrow X_1 +2\pi R_1$. Note 
that the above equations
also show that the $Z_2$ projection acts trivially on the instanton
position $a_{\mu}$ and scale $\rho$ as expected.  

The above result has also been obtained in \cite{horava}, where the
possibility to have orthogonal projections on $p$ and $p+4$ branes 
simultaneouly has been observed. Indeed by making T-duality along
the common world volume direction, we end up with IIA on $S^1$ modded
by $\Omega$ times the reflection on $S^1$. The orientifold planes at
the two fixed points come with opposite sign ${\cal{O}}_8^+$ and
${\cal{O}}_8^-$. The D4-branes and
D0-branes come with symplectic and orthogonal projections respectively
at ${\cal{O}}_8^-$ and vice versa at ${\cal{O}}_8^+$. Thus taking D4-branes
at one fixed point and D0-branes at other fixed point we can obtain 
orthogonal groups in both the systems. The fact that in order for this
to happen, the D4-brane and D0-brane systems
must sit at the two different fixed points implies that in the T-dual
description there is a relative $Z_2$ Wilson line between the D5-branes
and D1-branes which makes the bifundamental fields anti-periodic.

\section{Partition functions and symmetric product spaces}
           
In this section we derive a general formula for the character valued
string partition function of a symmetric product
CFT, involving fields carrying non-trivial boundary conditions.
More precisely, we will consider the orbifold CFT defined as
the symmetric product $S_N {\cal H}\equiv {\cal H}^N/S_N$,  
with ${\cal H}$ describing the Hilbert 
space of closed string excitations,
with periodic or antiperiodic boundary
conditions along $\sigma$ and $\tau$ 
directions of the worldsheet torus, 
which we will denote by $(\sigma_1,\sigma_2)$.
We label the boundary conditions by the 
characteristics ${g_\phi \brack h_\phi}$, with 
$g_\phi,h_\phi$ taking values $\{g_\phi,h_\phi=0,{1\over 2}\}$, 
describing the holonomies of a given field $\Phi$ around the two cycles: 
\be
\Phi(\sigma_1+1,\sigma_2)=e^{2\pi i g_\phi}\Phi(\sigma_1,\sigma_2)\quad\quad
\Phi(\sigma_1,\sigma_2+1)=e^{2\pi i h_\phi}\Phi(\sigma_1,\sigma_2)
\label{bcphi}
\ee
Our results generalize the familiar symmetric product formula  
\cite{dmvv} to the case where some of the fields carry 
boundary conditions different from the periodic  ones
($g_\phi=h_\phi=0$ in our notation) and
can be associated to sectors of a diagonal $Z_2$ orbifold 
action on 
the more familiar symmetric product spaces. 
As we have seen in the previous section, such CFTs 
naturally arise in the study of D-brane bound
state physics for type IIB orbifolds/orientifolds 
involving $Z_2$-shifts in the winding-momentum modes.

The derivation of the partition function follows, with slight
modifications, the procedure of \cite{dmvv} 
(see also \cite{bgmn}).
We start by specifying the character valued string partition 
function for single copy of the Hilbert space ${\cal H}$ 
\bea
{\cal Z}{g\brack h}({\cal H}|q,\bar{q},y)&=&
{\rm Tr}_{{\cal H}}\, q^{L_0-c/24}\,\bar{q}^{\bar{L}_0-c/24}
\,y^{J^3_0} \,\tilde{y}^{\bar{J}^3_0}\nonumber\\
&=&\sum
\,C{g\brack h}(\Delta,\bar{\Delta},\ell,\tilde{\ell})
\,q^{\Delta}\,\bar{q}^{\bar{\Delta}}\, y^\ell \, \tilde{y}^{\tilde{\ell}}
\label{zh}
\eea
with the sum running over $\Delta,\bar{\Delta},\ell,\tilde{\ell}$. 
The supertrace runs over string states in the Hilbert space
${\cal H}$ and 
boundary conditions for the various sigma model fields
are compactly denoted by ${g\brack h}$,
$q=e^{2\pi i\tau}$ describes the genus-one 
worldsheet modulus, $L_0$, $\bar{L}_0$ are the Virasoro generators.
$J_0^3,\bar{J}_0^3$ are the Cartan generators  of a  
$SU(2)_L\times SU(2)_R$ 
current algebra,  present in all our models,
to which $y$ and $\tilde{y}$ couple respectively.

Our next task is to evaluate the supertrace (\ref{zh}) for
a Hilbert space constructed by considering $N$ copies of the Hilbert
space ${\cal H}$ modded out by the permutation group $S_N$. 
This can be done following the standard non-abelian orbifold techniques 
developed in \cite{dhvw}. The string partition function is
written as the double sum  
\be
Z_{gh}=\sum_{gh=hg}\,{1\over C_{g,h}}\,\, [h] 
\Box \raisebox{-12pt}{\hspace*{-12pt} $[g]$} 
\label{zgh}
\ee
over orbifold twisted sectors labeled by the 
conjugacy classes $[g]$ of the permutation group $S_N$
\be
[g]=\prod_{L=1}^n (L)^{N_L} \quad {\rm with}\quad  \sum_{L=1}^n L N_L=N,
\label{g0}
\ee 
and over the conjugacy classes $[h]$ of the centralizer
\be
{\cal C}_g = \prod_{L=1}^n {\cal S}_{N_L}\ltimes Z_L^{N_L}
\ee
The integer $C_{g,h}$ is the order of the centralizer of $h$ in $\cal{C}_g$.
In writing (\ref{zgh}) we have
used the fact that traces of elements in the same conjugacy class
$[h]$ lead to the same result. 
We will write elements in $[h]$ as
\bea
[h]=\prod_{L=1}^n \prod_{M=1}^{n_L}\, (M)^{r_M^L}\, {\bf t}\,   
\in \prod_{L=1}^{n} {\cal S}_{N_L}\ltimes Z_L^{N_L}
\quad \sum_{M=1}^{n_L} M r_M^L=N_L.
\label{h}
\eea
with ${\bf t}$ an element in ${\bf Z_L}^{N_L}$ for a given choice of $N_L$'s. 

Orbifold group sectors are then parametrized by the integers $\{N_L\}$
(partitions of N), $\{r_M^L\}$ (partitions of $N_L$) 
and $\{ {\bf t}\}$ (elements of $Z_L^{N_L}$). In 
particular the number of such $[h]$'s is given by 
\be
{\cal C}_{g,h}=\prod_{L,M} L^{r_M^L} M^{r_M^L} r_M^L !
\ee
We can label the $N$ copies of the field in $S_N{\cal H}$ by the quintuple
of integers $(L,l,M,m,i)$
running over the domains $L=1,\dots,n$, $l=0,1,\dots, L-1$
$M=1,\dots, n_L$, $m=0,1,\dots, M-1$ and $i=1,\dots, r_M^L$
respectively. Writing $Z_L^N$ elements as ${\bf t}=(L)^{s_{m;i}}$
the boundary conditions for a field $\Phi_l^{m;i}$
(dependence in $L,M$ are implicitly understood) along the 
worldsheet torus cycles can be written as
\bea
\Phi_l^{m;i}(\sigma_1+1,\sigma_2)&=&e^{2\pi i g_\phi}
\Phi_{l+1}^{m;i}(\sigma_1,\sigma_2)\nonumber\\
\Phi_l^{m;i}(\sigma_1,\sigma_2+1)&=&e^{2\pi i h_\phi}
\Phi_{l+s_{m;i}}^{m+1;i}(\sigma_1,\sigma_2)
\label{bcn}
\eea
After iterating (\ref{bcn}) one is left with the doubly periodic 
functions
\bea
\Phi_l^{m;i}(\sigma_1+L,\sigma_2)&=&e^{2\pi i L g_\phi }
\Phi_{l}^{m;i}(\sigma_1,\sigma_2)\nonumber\\
\Phi_l^{m;i}(\sigma_1+s_i,\sigma_2+M)&=&e^{2\pi i (s g_\phi+M h_\phi)}
\Phi_{l}^{m;i}(\sigma_1,\sigma_2)
\label{bcnf}
\eea 
with $s_i=\sum_{m=0}^{M-1}\, s_{m;i}~({\rm mod}~L)$.
The contribution to the orbifold string partition function
of a given sector specified by $\{ N_L, r_M^L, s_{m;i} \}$
can therefore be written
in terms of the result for a single copy in a torus with 
the induced complex structure $\tilde{\tau}={M\over L}\tau+{s\over L}$
and spin characteristics (\ref{bcnf}), i.e.
\bea
(M)^{r_M^L} {\bf t}\,\,
\square \raisebox{-12pt}{\hspace*{-12pt} $(L)^{N_L}$}: 
\prod_{i=1}^{r_M^L} 
Z{g L \brack g s_i+M h} (\tilde{q}_i,\tilde{\bar{q}}_i,y^M,
\tilde{y}^M)  
\label{bt}
\eea
with $\tilde{q}_i=e^{2\pi i \tilde{\tau}_i}=
q^{M\over L}e^{2\pi i {s_i\over L}}$

Plugging this basic trace result into the sum over orbifold sectors
specified by (\ref{g0}), (\ref{h}), we are left with  
\bea
&&{\cal Z}{g\brack h}(S_N M|q,\bar{q},y,\tilde{y})=\nonumber\\
&&\sum_{\{N_L\},\{r_M^L\},\{s_i^L\}}\prod_{L,M,i}
  {1\over M^{r_M^L} r_M^L !}{1\over L^{r_M^L}}
{\cal Z}{g L \brack g s_i^L+M h}(\tilde{q}_i,
\bar{\tilde{q}}_i, y^M,\tilde{y}^M)
\nonumber\\
&&=\sum_{\{N_L\},\{r_M^L\}}
\prod_{L,M}{1\over M^{r_M^L} r_M^L!}\nonumber\\
&&\times\left({1\over L} 
\sum_{s=0}^{L-1}
\,C{g L\brack g s+M h}
(\Delta,\bar{\Delta},\ell,\tilde{\ell})
\,q^{M\Delta\over L}\,\bar{q}^{M\bar{\Delta}\over L} 
\,e^{2\pi i {s\over L}(\Delta-\bar{\Delta})}
\,y^{M\ell} \,\tilde{y}^{M\tilde{\ell}} \right)^{r_M^L}
\label{genus1}
\eea
with $C{g \brack h}
(\Delta,\bar{\Delta},\ell,\tilde{\ell})$ the expansion coefficients
(\ref{zh}) defined for a single copy of ${\cal H}$.

Before proceeding further, it is worth to make a comment 
about the BPS content of this
formula. We have seen in the previous section that 
the CFTs  describing excitations of the D-brane bound state 
involve fermionic zero modes. The trace over these modes leads to the
vanishing of the quantity inside the bracket in (\ref{genus1}) for
$y=\tilde y=1$,
corresponding to the fact that bound state excitations organize
themselves into supermultiplets of the unbroken supersymmetry.   
Sectors with $r_M^L=1$ correspond then to states in  ultra-short 
BPS supermultiplets and the counting formula for these states
simplifies to \cite{bgmn}:
\bea
&&{\cal Z}_{BPS}{g\brack h}(S_N M|q,\bar{q},y,\tilde{y})=
\nonumber\\&&~~{1\over N} 
\sum_{s,L,M}
\,C{g L\brack g s+M h}
(\Delta,\bar{\Delta},\ell,\tilde{\ell})
\,q^{M\Delta\over L}\,\bar{q}^{M\bar{\Delta}\over L} 
\,e^{2\pi i {s\over L}(\Delta-\bar{\Delta})}
\,y^{M\ell}\, \tilde{y}^{M\tilde{\ell}} 
\label{genusbps}
\eea
with $N=LM$, $s=0,1,\dots, L-1$. 
In some of our considerations in the following section, the 
restriction of the general formula 
presented below to this sector will be relevant. 
           
Coming back to the general expression (\ref{genus1}) one can now
perform the sum over $s$. It is easy to
see (see the appendix A for similar projection-sum manipulations) 
that this leads effectively to a projection onto 
states satisfying the ``level matching condition''
\be
{(\Delta-\bar{\Delta})\over L}\in {\bf Z}+\delta.
\label{lm}
\ee 
with $\delta=0,1/2$ depending on the different orbifold group
sectors and boundary conditions. Introducing, as in \cite{dmvv}, 
a generating function for the symmetric product formulae (\ref{genus1}) 
\be
{\cal Z}{g\brack h}(p,q,\bar{q},y,\tilde{y})=\sum_{N\geq 0} p^N
{\cal Z}{g\brack h}(S_N {\cal H}|q,\bar{q},y,\tilde{y})
\ee
with $p^N=p^{LM r_M^L}$, one can
write the final result in the compact form:
\bea 
{\cal Z}{g\brack h}(p,q,\bar{q},y,\tilde{y})=
\prod_{\delta=0}
(1-p^L q^{\Delta\over L}\bar{q}^{\bar{\Delta}\over L} y^{\ell}
\tilde{y}^{\tilde{\ell}})^{-C_+\left\{gL\right\}
(\Delta,\bar{\Delta},\ell,
\tilde{\ell})} &&\nonumber\\
\times \prod_{\delta={g}}
(1-(-)^{2 h} p^L q^{\Delta\over L}\bar{q}^{\bar{\Delta}\over L} y^{\ell}
\tilde{y}^{\tilde{\ell}})^{-C_-\left\{gL\right\}
(\Delta,\bar{\Delta},\ell,
\tilde{\ell})}.&& 
\label{genusdd}
\eea
The products run over all possible 
$L,\Delta,\bar{\Delta},\ell,\tilde{\ell}$ 
satisfying the level matching condition (\ref{lm}) with $\delta$
explicitly indicated in (\ref{genusdd}). The coefficients 
$C_\pm\left\{g\right\}$
are defined by 
\be
C_\pm\left\{g\right\}={1\over 2}\left(C{g\brack 0}\pm C{g\brack {1\over 2}}
\right)
\label{cpm}
\ee
and count the number of states in 
the $g$-twisted sector of the starting (the single copy)
CFT with $\pm$ eigenvalues under the $Z_2$ orbifold group action.
The choice $g=h=0$ corresponds
to the case studied in \cite{dmvv} where all fields are 
periodic in both $\sigma$ and $\tau$ directions and the sum
over $s$ results in a projector onto (\ref{lm}) with 
$\delta=0$. 

Specifying to ground states in the right moving
sector of the symmetric product CFT, the formula (\ref{genusdd}) 
reduces to 
\bea 
{\cal Z}{0\brack 0}(p,q,y,\tilde{y})&=&
\prod_{L,k}
(1-p^L q^k y^{\ell}\tilde{y}^{\tilde{\ell}})^{-C{0\brack 0}(Lk,\ell,
\tilde{\ell})}
\nonumber\\
{\cal Z}{0\brack {1\over 2}}(p,q,y,\tilde{y})&=&
\prod_{L,k}
(1-p^L q^k y^{\ell}\tilde{y}^{\tilde{\ell}})^{-C_+ \{ 0 \}}
(1+p^L q^k y^{\ell}\tilde{y}^{\tilde{\ell}})^{-C_-\{ 0 \}}
\label{11}\\
{\cal Z}{{1\over 2}\brack 0}(p,q,y,\tilde{y})&=&
\prod_{L,k}
(1-p^{2L} q^k y^{\ell}\tilde{y}^{\tilde{\ell}})^{-C_+\{ 0 \}}
(1-p^{2L-1} q^k y^{\ell}\tilde{y}^{\tilde{\ell}})^{-C_+\{ {1\over 2}\}}
\nonumber\\
&&\times (1-p^{2L} q^{k-{1\over 2}} y^{\ell}\tilde{y}^{\tilde{\ell}})^
{-C_-\{ 0 \}}
(1-p^{2L-1} q^{k-{1\over 2}} y^{\ell}\tilde{y}^{\tilde{\ell}})^{-C_- 
\{{1\over 2}\}}
\nonumber\\
{\cal Z}{{1\over 2}\brack {1\over 2}}(p,q,y,\tilde{y})&=&
\prod_{L,k}
(1-p^{2L} q^k y^{\ell}\tilde{y}^{\tilde{\ell}})^{-C_+\{ 0 \}}
(1-p^{2L-1} q^k y^{\ell}\tilde{y}^{\tilde{\ell}})^{-C_+\{ {1\over 2}\}}\nonumber\\
&&\times (1+p^{2L} q^{k-{1\over 2}} y^{\ell}\tilde{y}^{\tilde{\ell}})^
{-C_-\{ 0 \}}
(1+p^{2L-1} q^{k-{1\over 2}} y^{\ell}\tilde{y}^{\tilde{\ell}})^{-C_-
\{{1\over 2}\}}
\nonumber
\eea
where the arguments of the expansion coefficients 
$C_{\pm}\{g\}(nm, \ell,\tilde{\ell})$, with $n$ and $m$ being the
powers of $p$ and $q$ respectively have been omitted.  
The net effect of a non-trivial holonomy $(g,h)\neq (0,0)$
is then to correlate the parity of excitations in the CFT  
under the $Z_2$ orbifold group action to the parity
of the permutation group orbifold sector and the level
of the SCFT specified by $L$ and $k\equiv {\Delta\over L}$
respectively\footnote{Partition functions somewhat similar to
(\ref{11}) appear in \cite{D2}, where however the symmetric product
orbifold includes discrete torsion}. 

Note that the marginal
deformations of the CFTs are given by the ground states in
the above partition functions which appear as $p^n
(y\tilde{y})^{1-n}$ as one can see by spectral flow from the Ramond
to the Neveu-Schwarz sector. One can then verify that for
model $I$ and $III$, the partition functions above  include the 
marginal deformations corresponding to the $T^4$ (they appear in the
untwisted sector of $S_N$) but they do not include the blowing up
modes which arise in the twisted sector of $S_N$. This is in
agreement with the discussion at the end of section 3.1.

\section{D1/D5 bound states versus fundamental strings}

In this section we evaluate the partition functions 
(or elliptic genera) encoding multiplicities
and charges of D1/D5 two-charge bound state systems and compare
them with the expected results from a U-dual description, 
in terms of winding-momentum modes of fundamental strings. 
We are assuming that the effective gauge theories, 
describing the low energy D1/D5 dynamics,
flow to one of the orbifold symmetric product 
CFTs (\ref{mc}), (\ref{mh}) with the $Z_2$ actions specified in the 
tables 2.1 and 2.2.
 
There is an important difference between the symmetric product
CFTs (\ref{mc}) and (\ref{mh}).
In the former case, associated to stacks of pure 
$D1(D5)$-branes bound to KK momentum modes, the position
in ${\bf R}^4$ is described by the center of mass of $N$ copies
of ${\bf R}^4$ in the symmetric product CFT. This leads to
a subtlety in the counting of BPS excitations, since not all
states contributing to the elliptic genus correspond to normalizable
ground states of the gauge theory \cite{gmnt}. A careful analysis
\cite{gmnt} reveals that among all the states with the right
supersymmetry structure to reconstruct a short supermultiplet ($r_M^L=1$)
in (\ref{genusbps}) only those coming from the long string sector
$[g]=(N)$ represent truly one-particle states. BPS charges 
and multiplicities can therefore be 
read off from the formula (\ref{genusbps})
specifying $M=1,L=N$, the so called ``long string''
sector. This is not the case for the symmetric products 
associated to D1/D5 bound states (\ref{mh}), where all intermediate
strings are needed in  (\ref{11})
in order to reproduce the fundamental string degeneracies.
The position of the bound state is, in this case, 
specified by a single coordinate in the non compact
transverse ${\bf R}^4$ (strictly speaking in our case ${\bf R}^4$
is replaced by ${\bf R}^3\times S^1$).

\subsection{Fundamental string partition functions}  
  
Before proceeding with the study of the spectrum of D-brane 
bound states, let us evaluate the partition function for the fundamental
sides of the duality chain. Two-charge D-brane bound
state will be systematically mapped to a fundamental strings
carrying both momentum ($p_1$) and winding charges ($F_1$), with 
no extra charges turned on in one of the three type IIB 
orbifolds corresponding to  $(-)^{F_L} I_4 \sigma_{p_a}$, 
$(-)^{F_L}\sigma_{p_a}$ and   $I_4\sigma_{p_a}$. We will refer to
these theories as $I_F$, $II_F$ and $III_F$ respectively. 
$\sigma_{p_a}$ represents a $Z_2$-shift in the momentum mode 
along direction $a$, with $a=1,6$ in the case of a longitudinal
and transverse shift respectively.

The fundamental string partition functions for the BPS
states is defined by the
supertrace (\ref{zh}), restricted to  the right-moving 
ground states.
After performing the spin structure sums, these can be written as
\bea
&&Z_{I_F}(q,y,\tilde{y})=
{1\over 2}\,{{y}_-^2\vartheta_1^2(y)\over 
\hat{\vartheta}_1(y \tilde{y})
\hat{\vartheta}_1(y \tilde{y}^{-1})}\left(\Gamma_{4,4}
\tilde{y}^2_-\,{\vartheta_1^2(\tilde{y})\over \eta^6}
\Gamma_{1,1}{0\brack 0} +\right.\nonumber\\
&&~~~\left. \tilde{y}_+^2\,{\vartheta_2^2(\tilde{y})\over \hat{\vartheta}_2^2(0)}
\Gamma_{1,1}{0\brack {1\over 2}}
+16 {\vartheta_4^2(\tilde{y})\over \hat{\vartheta}_4^2(0)}
\Gamma_{1,1}{{1\over 2}\brack 0}+
16 {\vartheta_3^2(\tilde{y})\over \hat{\vartheta}_3^2(0)}
\Gamma_{1,1}{{1\over 2}\brack {1\over 2}}\right)\label{if}\\
&& Z_{II_F}(q,y,\tilde{y})={1\over 2}\,{y_-^2 \tilde{y}_-^2 \over 
\hat{\vartheta}_1(y \tilde{y})
\hat{\vartheta}_1(y \tilde{y}^{-1})\eta^6}\,\Gamma_{4,4}
\left(\vartheta_1^2(y)\vartheta_1^2(\tilde{y})\Gamma_{1,1}{0\brack 0}
\right.\nonumber\\
&&~~~\left.+\vartheta_2^2(y)\vartheta_2^2(\tilde{y})
\Gamma_{1,1}{0\brack {1\over 2}}
 +\vartheta_4^2(y)\vartheta_4^2(\tilde{y})
\Gamma_{1,1}{{1\over 2}\brack 0}
+\vartheta_3^2(y)\vartheta_3^2(\tilde{y})
\Gamma_{1,1}{{1\over 2}\brack {1\over 2}}\right)\label{iif}\\
&& Z_{III_F}(q,y,\tilde{y}) ={1\over 2}\,{y_-^2\vartheta_1^2(\tilde{y})\over 
\hat{\vartheta}_1(y \tilde{y})
\hat{\vartheta}_1(y \tilde{y}^{-1})}\left(\Gamma_{4,4}
\tilde{y}_-^2\,{\vartheta_1^2(y)\over \eta^6}
\Gamma_{1,1}{0\brack 0}\right.\nonumber\\
&&~~~\left. +\tilde{y}_+^2\,{\vartheta_2^2(y)\over \hat{\vartheta}_2^2(0)}
\Gamma_{1,1}{0\brack {1\over 2}}
+16 {\vartheta_4^2(y)\over \hat{\vartheta}_4^2(0)}
\Gamma_{1,1}{{1\over 2}\brack 0}+
16 {\vartheta_3^2(y)\over \hat{\vartheta}_3^2(0)}
\Gamma_{1,1}{{1\over 2}\brack {1\over 2}}\right)\label{iiif}
\eea

with $y_{\pm}\equiv y^{1\over 2}\pm y^{-{1\over 2}}$ and similarly
for $\tilde{y}_{\pm}$. 
$\Gamma_{4,4}$ is the $T^4$ winding-momentum lattice sum and 
\be
\Gamma_{1,1}{g \brack h}\equiv
\sum_{(p_a,w_a)\in ({\bf Z},{\bf Z}+g)}
\, (-)^{2 p_a h} q^{(p_a/R+w_a R)^2}\, \bar{q}^{(p_a/R-w_a R)^2}.
\label{latt11}
\ee
is the shifted lattice parallel to $\sigma_{p_a}$. 
The hat in the $\vartheta$-functions in the denominators
denotes the omission of their zero mode parts, i.e. 
$\hat{\vartheta}_1\equiv \eta^3$, 
$\hat{\vartheta}_2\equiv {1\over 2}\vartheta_2$ and 
$\hat{\vartheta}_{3,4}\equiv \vartheta_{3,4}$.  
The completely untwisted sector, common to all three models, corresponds
to the partition function of type IIB on $T^5$. 
In the case of a transverse shift, an extra $\Gamma_{1,1}{0\brack 0}$
lattice sum, common to all orbifold group sectors, should be included.

Multiplicities for fundamental string states, carrying $k$ units 
of momenta and $N$ units of winding, can be read off 
from (\ref{if})-(\ref{iiif}),
once the level matching condition ($N_R=c_R$),
\be
kN=N_L-c_L~,
\label{lmf}
\ee
is imposed. 
Here $N_L,N_R$ are the left- and right-moving oscillator levels 
and $c_L,c_R$ the zero point energies.

The fourth fundamental theory that will be relevant to our next 
discussion is the toroidal heterotic string  
with gauge group $SO(32)$ completely broken by Wilson lines.
A possible choice of Wilson lines (in a fermionic representation)
can be written as
\bea
A_1 &:& \left[\, (+)^{16}, (-)^{16} \, \right]\nonumber\\
A_2 &:& \left[\, (+)^{8}, (-)^8, (+)^{8}, (-)^8\, \right]\nonumber\\    
    &\cdot&\nonumber\\
    &\cdot&\nonumber\\
A_5 &:& \left[\, (+), (-), (+),(-), .... (+), (-)  \,\right]
\label{wl}
\eea
Alternatively, one can represent this model as a $Z_2^5$ orbifold 
of heterotic string on $T^5$, where the $Z_2$ generators act simultaneously
as a shift in one of the five circles and on the $SO(32)$ lattice,
in the way specified by (\ref{wl}).
The fundamental string partition function can then be written as
\bea
Z_{IV}(q,y,\tilde{y})&=&{1\over 2^5}\, {y_-^2 \tilde{y}^2_- \over 
\hat{\vartheta}_1(y \tilde{y})
\hat{\vartheta}_1(y \tilde{y}^{-1})
\eta^{18}} 
\left( {1\over 2}(\vartheta_2^{16}+\vartheta_3^{16}+\vartheta_4^{16})
\,\Gamma_{5,5}{0\brack 0}\right.\nonumber\\
&& \left. 
+\vartheta_3^{8}\vartheta_4^{8}\,\Gamma_{5,5}{0\brack \epsilon_i}
+\vartheta_4^{8}\vartheta_2^{8}\,\Gamma_{5,5}{0\brack \epsilon_i}
+\vartheta_3^{8}\vartheta_2^{8}\,\Gamma_{5,5}{\epsilon_i\brack \epsilon_i}
\right)
\label{i4}
\eea
where the sum over $\epsilon_i=0,{1\over 2}$ with 
$i=1,\dots,5$ is always implicitly understood. 
We denote by $\Gamma_{5,5}{g_i\brack h_i}$ the lattice built 
from five copies of (\ref{latt11}), with twists specified 
by $g_i,h_i$. Notice that among all
the $Z_2^5$ orbifold group elements only the $\epsilon_i$-projection
(for a fixed $\epsilon_i$) leads to a non-trivial result in 
the $\epsilon_i$-twisted sector.

\subsection{D1(D5)-KK momenta bound states}

In this subsection we compare the CFT results (in the long string sector) 
for multiplicities of BPS states,
i.e. $\bar{\Delta}=0$, in the pure D1(D5)-KK bound state systems, 
to the ones coming from the fundamental string partition functions
(\ref{if})-(\ref{iiif}) in the dual theories. The partition function will be
evaluated using the CFT proposals (\ref{mc}) with $Z_2$ given in table 2.1. 
As explained before, only the long string sector, $M=1, L=N$ 
in (\ref{genusbps})
is relevant to the counting of
one-particle states. The results
generalize a similar analysis in \cite{bgmn}.

In the transverse shift case, the CFT description
of a pure D1-KK or D5-KK systems is associated to the 
untwisted sector of  $Z_2$ orbifolds 
of $(R^3\times S^1\times T^4)^N/S_N$. 
Specializing to the long string sector in (\ref{genusbps}), 
with $g=0,h={1\over 2}$, one is left with
\be
{1\over 2}\left({\cal Z}_{long}{0\brack 0}+{\cal Z}_{long}
{0\brack {1\over 2}}\right)
=\sum \,C_+ \{0 \}(kN,\ell,\tilde{\ell})\,
q^k\, y^\ell\, \tilde{y}^{\tilde{\ell}}
\ee
where $C_{\pm} \{g \}(kN,\ell,\tilde{\ell})$ are the expansion 
coefficients of the partition function evaluated for a single 
copy $N=1$ in (\ref{mc}). 
We have performed the sum over $s=0,1\dots, N-1$, that projects
the sum onto states satisfying the level matching condition
$k={\Delta \over N}\in {\bf Z}$.        
Charges and multiplicites for a bound state of $N$ 
D1- or D5-branes carrying $k$ units of KK 
momentum $p_1$ are then described 
in each theory by the corresponding expansion coefficients 
$C_+\{0 \}(kN,\ell,\tilde{\ell})$. Recalling from our discussion
in section 2, that the 
$N=1$ CFTs, and therefore their $C_+\{0 \}$ coefficients
in table 2.1, coincide in each of the cases
with their fundamental descriptions, 
we conclude that the D1(D5)-KK
multiplicities we obtain from the above 
CFTs agree with those  of   
untwisted states (even windings) with even momenta, 
in the corresponding fundamental theory (\ref{if})-(\ref{iiif}),
once the level matching condition (\ref{lmf}) is imposed. 
This is precisely what one would expect from the duality map, since 
the image of the D1(D5)-KK  bound state ($p_6=F_6=0$)
carries no windings and no momenta along the shift. 
      
A similar result can be found in the longitudinal shift case $\sigma_{p_1}$.
Now the long string sector in (\ref{genusbps}), with $g={1\over 2},h=0$
leads to 
\be
{\cal Z}_{long}{{1\over 2}\brack 0}=\sum_{k\in {\bf Z}} 
\,C_+ \{ {N\over 2} \}(kN,\ell,\tilde{\ell})\,
q^k\, y^\ell\, \tilde{y}^{\tilde{\ell}}
+\sum_{k\in {\bf Z}+{1\over 2}} 
C_- \{ {N\over 2} \}(kN,\ell,\tilde{\ell})\,
q^k\, y^\ell\, \tilde{y}^{\tilde{\ell}}
\ee
This is again in complete agreement with (\ref{if})-(\ref{iiif}). 
Even(odd) fundamental winding states are mapped to bound states
involving an even(odd) number N of D-branes and multiplicities
are described by $C_{\pm}$
according to whether the level $k$ (momentum on the fundamental side)
is integer or half-integer. 

\subsection{D1/D5 bound states}

In this subsection we consider D1/D5 bound state systems. In order to
compare multiplicities of the bound states with 
those of fundamental strings (pure $F_1-p_1$), 
one should restrict the
attention to ground states $\Delta=\bar{\Delta}=0$ in both left-
and right-moving sides of the CFT. 
Once again, the transverse shift results can be expressed as $Z_2$
orbifolds:    
\be
Z(p,y,\tilde{y})=\sum_{h=0,{1\over 2}} Z_{cm}{0\brack h}(0,y,\tilde{y})
\, Z_{sym}{0\brack h}(p,y,\tilde{y})
\ee
of the type IIB result  
\be
Z_{cm}{0\brack 0}(0,y,\tilde{y})\, Z_{sym}{0\brack 0}(p,0,y,\tilde{y})
=y^2_- \tilde{y}^2_- \, \frac{\vartheta_1^2(y|p)
\vartheta_1^2(\tilde{y}|p)}
{\hat{\vartheta}_1(y \tilde{y}|p)
\hat{\vartheta}_1(y \tilde{y}^{-1}|p)\eta^6(p)}.
\label{iib}
\ee
associated to 
the symmetric product space $R^3\times S^1\times T^4\times (T^4)^N/S_N$.
We denote by $Z_{cm}(q,y,\tilde{y})$ the contribution 
coming from the center of mass,
while $Z_{sym}(p,q,y,\tilde{y})$ will be associated 
to the symmetric product of the
$T^4$ tori. Using the data for the single copy the CFT $T^4/\hat{g}$
with $\hat{g}$ specified in table 2.2:
\bea
(-)^{F_L}: && \sum_{\ell,\tilde{\ell}}\, 
C^{I}_{\pm}\{0\}(0,0,\ell,\tilde{\ell})\,y^\ell\, \tilde{y}^{\tilde{\ell}}
={1\over 2}\, (\tilde{y}^2_-\,y_-^2\pm\tilde{y}^2_-\, y_+^2)\nonumber\\
I_4^{sp}: && \sum_{\ell,\tilde{\ell}}\, 
C^{II}_{\pm}\{0\}(0,0,\ell,\tilde{\ell})\,y^\ell\, \tilde{y}^{\tilde{\ell}}
={1\over 2}\, (\tilde{y}^2_-\,y_-^2\pm\tilde{y}^2_+\, y_+^2)\nonumber\\
(-)^{F_L}\, I_4^{sp}: && \sum_{\ell,\tilde{\ell}}\, 
C^{III}_{\pm}\{0\}(0,0,\ell,\tilde{\ell})\,y^\ell\, \tilde{y}^{\tilde{\ell}}
={1\over 2}\, (\tilde{y}^2_-\,y_-^2\pm\tilde{y}^2_+\, y_-^2)
\eea
into the symmetric product formulae (\ref{11}), we are left with the partition 
functions 
\bea
&& Z_{I}{0 \brack {1\over 2}}(p,y,\tilde{y}) =
 {1\over 2}\ y_-^2\tilde{y}_+^2 \,{ \vartheta_2^2(\tilde{y}|p)\over 
\hat{\vartheta}_1(y \tilde{y}|p)
\hat{\vartheta}_1(y \tilde{y}^{-1}|p)}
\,{\vartheta_1^2(y|p)\over \hat{\vartheta}_2^2(0|p)}\nonumber\\
&& Z_{II}{0 \brack {1\over 2}}
(p,y,\tilde{y})={1\over 2}\, y_-^2\tilde{y}_-^2 \,{ 
\vartheta_2^2(\tilde{y}|p)\over 
\hat{\vartheta}_1(y \tilde{y}|p)
\hat{\vartheta}_1(y \tilde{y}^{-1}|p)}
\,{\vartheta_2^2(y|p)\over \eta^6(p)}\nonumber\\
&& Z_{III}{0 \brack {1\over 2}}(p,y,\tilde{y}) =
 {1\over 2}\,y_-^2\tilde{y}_+^2 \, {\vartheta_1^2(\tilde{y}|p)\over 
\hat{\vartheta}_1(y \tilde{y}|p)
\hat{\vartheta}_1(y \tilde{y}^{-1}|p)}
\,{\vartheta_2^2(y|p)\over \hat{\vartheta}_2^2(0|p)}\label{d1d5t}
\eea

where the contributions of the center of mass 
$Z^I_{cm}(y,\tilde{y})=y_-^4 \tilde{y}^4_+$,
 $Z^{II}_{cm}(y,\tilde{y})=Z^{III}_{cm}(y,\tilde{y})=
y_-^2 y_+^2 \tilde{y}^2_- \tilde{y}_+^2$ in the 
three cases, has been included.

Together with (\ref{iib}) the  results (\ref{d1d5t})
for D1/D5 bound state degeneracies
reproduce the multiplicities (\ref{if})-(\ref{iiif}) 
for untwisted fundamental strings with $p_6=0$, as required by 
the U-duality chain of table 1.2. 

Finally, let us compute the D1/D5 bound state spectrum on
$T^4\times S^1/I_4\sigma_{p_1}$ with $\sigma_{p_1}$ a longitudinal shift.
The relevant CFT data are now given in terms of the expansion 
coefficients for $T^4/I_4^{sp}$:
\bea
\sum_{\ell,\tilde{\ell}}\, 
C^{IV}_{\pm}\{ 0 \}(0,0,\ell,\tilde{\ell})\,y^\ell\, \tilde{y}^{\tilde{\ell}}
&=&{1\over 2}\, (\tilde{y}^2_-\,y_-^2\pm\tilde{y}^2_+\, y_+^2)\nonumber\\
\sum_{\ell,\tilde{\ell}}\, 
C^{IV}_{+}\{ {1\over 2} \}(0,0,\ell,\tilde{\ell})\,y^\ell\, \tilde{y}^
{\tilde{\ell}}
&=& 16 \nonumber\\
\sum_{\ell,\tilde{\ell}}\, 
C^{IV}_{-}\{ {1\over 2} \} (0,0,\ell,\tilde{\ell})\,y^\ell\, 
\tilde{y}^{\tilde{\ell}}&=&0
\eea
and $Z_{cm}=16\, y_-^2 \tilde{y}_-^2$. Plugging in (\ref{11}) we are left
with
\bea
Z_{II}{{1\over 2} \brack 0}(p,y,\tilde{y})&=& 16\,
y_-^2 \tilde{y}_-^2\, { 1\over
\hat{\vartheta}_1(y \tilde{y})
\hat{\vartheta}_1(y \tilde{y}^{-1})
\eta^{2}}{\eta^8\over \vartheta_4^8(0)}\nonumber\\
&=&{1\over 2^4} y_-^2 \tilde{y}_-^2\, {1\over
\hat{\vartheta}_1(y \tilde{y})
\hat{\vartheta}_1(y \tilde{y}^{-1})
\eta^{18}}\, \vartheta_2^8(0) \vartheta_3^8(0)
\label{d1d5l}
\eea
This is in complete agreement with the fundamental heterotic string
degeneracies (\ref{i4}), coming from the twisted sector, once the
level matching condition (\ref{lmf}) is imposed. 
Notice that the expansion of (\ref{d1d5l})
reproduces both signs in (\ref{i4}), $p_1$ even or odd, according to
whether we expand to integer or half-integer powers in $p$. 
That only states in the
twisted sector (odd windings) are relevant to the comparison is due to the
fact that the CFT proposals are valid only for a single fivebrane,
which is mapped in the fundamental side to a single unit of winding mode.
One can however test multiplicities in the untwisted sector (with $p_1=1$)
by going after four T-dualities to step {\bf B}, where the role of
the D1 and D5 are exchanged inside model $II$. The CFT description of those
D1/D5 states are, of course, the same as before and multiplicities
are again given by (\ref{d1d5l}). The fundamental string multiplicities
on the other hand lead to apparently two very different results,
depending on whether we consider states with $F_1$ odd (twisted sector)
or even (untwisted sector). In the former case one finds again (\ref{d1d5l})
in agreement with the duality predictions. 
The multiplicities for even $F_1$  
are, on the other hand, given by  
\be
 Z^{p_1=1}_{F_1-even}(q,y,\tilde{y})={1\over 2^5}\, {y_-^2 \tilde{y}^2_- \over 
\hat{\vartheta}_1(y \tilde{y})
\hat{\vartheta}_1(y \tilde{y}^{-1})
\eta^{18}} 
\left[ {1\over 2}(\vartheta_2^{16}+\vartheta_3^{16}+\vartheta_4^{16})
-\vartheta_3^{8}\vartheta_4^{8}\right]
\label{zeven}  
\ee           
The discrepancy is only apparent, since 
expression (\ref{zeven}) coincides with (\ref{d1d5l})
after simple manipulations of $\vartheta$-identities. The fact that the 
heterotic dual model treats on the same footing winding and momentum
modes, as required by the U-duality chain, can
be taken as a further support for the consistency of the whole picture.

One can try to apply a similar analysis to the D1/D5 systems in models
$I$ and $III$ with a longitudinal shift, but one immediately runs
into problems. The ground states of the natural CFT proposals in table 2.2 
are in these cases either tachyonic or massive and a naive application
of the elliptic genus formula leads to meaningless results. 
A proper description of these (4,0) D1/D5 systems remains an
interesting open problem. 

\subsection{Three-charge systems}
We will restrict the discussion of 3-charge systems
to the transverse shift case, since, as we mentioned before,
the D1/D5 CFT description of models $I$ and $III$ in the longitudinal
shift case is problematic due to the presence of either tachyonic
or massive ground states.

To extract the multiplicities for 3-charge systems, i.e.
D1/D5 system carrying $k$ units of KK momentum, from
our partition functions, we restrict the right-moving
part to the ground state ($\bar\Delta=0$) and 
consider excitations of the left-moving part (which is 
non-supersymmetric in models $I$
and $III$) to level $k$. The resulting partition 
functions are of the form: 
\be
Z(p,q,y,\tilde{y})=\sum_{h=0,{1\over 2}} Z_{cm}{0\brack h}(q,y,\tilde{y})
\, Z_{sym}{0\brack h}(p,q,y,\tilde{y})
\label{zpq}
\ee
Notice that the $q^0$ term in $Z$ corresponds to
the partition function of the fundamental sides
for the D1/D5 systems and  is given in (\ref{d1d5t})
for the three theories. Denoting this latter  by $Z_F(p)$, it is
convenient to rewrite $Z$ in (\ref{zpq}) as:   
\be
Z(p,q,y,\tilde{y})=\sum_{h=0,{1\over 2}} {Z}_{cm}{0\brack h}(q,y,\tilde{y})
\,\hat{Z}_F{0\brack h}(p,y,\tilde{y}) 
\hat {Z}_{sym}{0\brack h}(p,q,y,\tilde{y})
\label{zpqf}
\ee
where, as before, the ``hat'' denotes omission of zero modes.

We have already seen in section 2 that U-duality
gives certain relations for the multiplicities of the 3 charge systems.
From table 1.2, by comparing columns {\bf F} and
{\bf G}, we see that models $I$ and $III$ get exchanged 
together with D1 and D5 charges. The CFTs we have proposed 
trivially satisfy this symmetry for $Q_5=1$, since in this case $N=1$
in (\ref{mh}) and therefore both CFTs are given by
$(R^3\times S^1\times T^4\times T^4/I_4)/(-)^{F_L}$. 

A less trivial relation comes from the comparison between
columns {\bf C} and {\bf F}: in this case the 3-charge system
(D1, D5, KK) of model $II$ is mapped to (KK, D5, D1) in model $III$. 
This means
that the full elliptic genera corresponding to 
models $II$ and $III$ must get exchanged if   
we exchange $q$ with $p$. 
Since:
\bea
\hat{Z}^{II}_{cm}{0\brack h}(q,y,\tilde{y})=
\hat{Z}^{III}_{F}{0\brack h}(q,y,\tilde{y})={1\over 2}\,
{{\hat \vartheta}_1^2(\tilde{y}|q)\over 
\hat{\vartheta}_1(y \tilde{y}|q)
\hat{\vartheta}_1(y \tilde{y}^{-1}|q)}
\,{{\hat\vartheta}_2^2(y|q)\over \hat{\vartheta}_2^2(0|q)}
\nonumber\\
\hat{Z}^{III}_{cm}{0\brack h}(q,y,\tilde{y})=
\hat{Z}^{II}_{F}{0\brack h}(q,y,\tilde{y})={1\over 2}\, 
{{\hat\vartheta}_2^2(\tilde{y}|q)\over 
\hat{\vartheta}_1(y \tilde{y}|q)
\hat{\vartheta}_1(y \tilde{y}^{-1}|q)}
\,{{\hat\vartheta}_2^2(y|q)\over \eta^6(q)}
\label{zII}
\eea
the symmetry under the $(p,q)$ exchange implies then that 
$\hat{Z}_{sym}{0\brack h}(p,q,y,\tilde{y})$ of model $II$ and $III$
should get exchanged.  
Notice that we are comparing
a $(4,4)$ theory (model $II$) with a $(4,0)$ theory (model
III). 
Although in the previous discussions we have
set $\bar\Delta=0$ while keeping 
$y,\tilde y$ arbitrary, actually the quantity that 
is expected to be invariant under deformations of the CFT (the
elliptic genus) is obtained
by setting ${\tilde y}=1$. However, in order to soak
the fermionic zero modes we will take two derivatives in $\tilde y$
and then set ${\tilde y}=1$ (the two derivatives necessarily act on
the center of mass CFT). This results in
putting the right-moving 
sector (which is supersymmetric in both cases) on the
ground state. Once we set $\tilde{y}=1$, the relevant expressions 
become (omitting indices $II$ and $III$):  
\bea
\hat{Z}_{sym}{0 \brack {1 \over 2}}(p,q,y,\tilde{y}=1)=\prod_{n,m \geq 1}
{\left(1+p^n q^m y^l \over 1-p^n q^m y^l\right)}^{{1\over 2}
C{0 \brack {1 \over 2}}(nm,l)}\,.
\label{zi}
\eea
(\ref{zi}) follows from (\ref{11}), using 
the identities:  
\bea
C{0 \brack {1 \over 2}}(m,l)\equiv \sum_{\tilde l} C_+\{0\}(m,l,\tilde{l})=
-\sum_{\tilde l} C_-\{0\}(m,l,\tilde{l})
\,.
\nonumber
\eea

The U-duality requirement that, under $(p,q)$ 
exchange $\hat{Z}_{sym}{0 \brack {1 \over 2}}$ for models $II$ and
$III$ 
are exchanged, implies 
\be
C_{II}{0 \brack {1 \over 2}}(m,l)=C_{III}{0 \brack {1 \over 2}}(m,l)\,, 
~~~~{\rm for}~~ m \geq 1. 
\label{CC}
\ee

This requirement is
not satisfied by the proposed CFTs for theories $II$ and $III$.
Indeed in terms of theta functions:
\bea
\sum_{m,l}C_{II}{0 \brack {1 \over 2}}(m,l)\, q^m\, y^l &=&
4{{\vartheta}_2^2(y|q) \over \hat{\vartheta}_2^2(0|q)}\,,
\nonumber\\
\sum_{m,l}C_{III}{0 \brack {1 \over 2}}(m,l)\, q^m\, y^l &=&
4{{\vartheta}_1^2(y|q) \over \hat{\vartheta}_2^2(0|q)}\,.
\eea
from which it follows that $C_{II}{0 \brack {1 \over 2}}(m,l)
\neq C_{III}{0 \brack {1 \over 2}}(m,l)$. Remarkably however the
equality
holds for $m$
odd !

This problem affects also the seemingly well understood case of 
the $(4,4)$ CFT ${\cal R}^4\times(K3)^N/S_N$, describing the 
D1/D5 system in type IIB on $K3\times S^1$.  
Using type II/heterotic duality in 6 dimensions,
one can relate type IIB on $K3\times S^1$ to  type IIB on 
$S^1\times T^4/\Omega I_4$, while exchanging D1 and KK charges.
Thus for D5 charge 1, this amounts to exchange $(p,q)$ 
(at order ${\bar q}^0$)
in the corresponding elliptic genus. Notice that in this case $Z_F(p)$ is 
just the bosonic oscillator part of the heterotic string and 
clearly ${\hat Z}_{sym}(p,q)$ is symmetric under
$(p,q)$ exchange \cite{dmvv, dvv, mms}. However it is also
easy to see that, for instance, the coefficient at order $q^1$
of the elliptic genus does not have a well defined
modular property as a function of $p$.
Finally, the same problem is present in the longitudinal 
shift case for the model II that we have studied before,
although ${\hat Z}(p,q)$ is not $(p,q)$ symmetric in this
case. 

In the above discussion we have ignored the non-trivial
background of RR 0- and 4-form fields in the symmetric product CFT. 
As pointed out in section 2, in the presence of this background   
the issue of $p$, $q$ exchange symmetry is more subtle. We will make
some more comments on this problem in the conclusion.

\section{One-loop effective gauge couplings}

In this section we study the $(T,U)$ moduli dependence of 
one-loop threshold corrections to ${\cal F}^{2k+4}$ gauge 
couplings in the low energy effective action, associated to
the four dimensional string compactifications under consideration. 
$(T,U)$ are
the Kahler and complex structure moduli of a $T^2$ along directions
{1,6}. The aim of this section 
is to extract this information for four-dimensional gauge couplings  
${\cal F}^{2k+4}$ involving some definite combinations 
of the eight field strengths,
\bea
{\cal F}^\pm_{Li}&\equiv&\partial_{[\mu}(G_{\nu] i}^\pm      
+ B_{\nu] i}^\pm     )\nonumber\\ 
{\cal F}^\pm_{Ri}&\equiv&\partial_{[\mu}(G_{\nu] i}^\pm 
- B_{\nu] i}^\pm     )
\label{flr}
\eea
which arise from KK-reductions of the six-dimensional
metric and antisymmetric tensor to $D=4$, with $i=1,6$.
$\pm$ refers to  (anti-)self-dual four-dimensional
two-forms. 
The one-loop threshold corrections for these couplings 
will be then mapped to
two-charge D-instanton contributions in the non-perturbative
U-dual descriptions of table 1.1.  
Related computations in various contexts can be found in \cite{thresholds}.
  
We will introduce a complex (euclidean) notation for spacetime 
coordinates    
\bea 
Z^1 &=& {1 \over \sqrt{2}}(X^0+i X^3) \quad
Z^2 = {1 \over \sqrt{2}}(X^1+i X^2) \quad
\nonumber\\
\chi^1 &=& {1 \over \sqrt{2}}(\psi^0+i \psi^3) \quad
~~\chi^2 = {1 \over \sqrt{2}}(\psi^1+i \psi^2) \quad
\eea
with barred quantities corresponding to  complex conjugates.

The moduli dependence will be extracted from the 
string amplitudes:
\be
{\cal A}_{2k+4}=\langle\,\prod_{i=1}^{k+2} 
V(p_1,\xi_i) V(\bar{p}_2,\tilde{\xi}_i)\rangle
\label{ak}
\ee
where for simplicity we choose a kynematical configuration where
half of the vertices carry momentum $p_1$ and the other half $\bar{p}_2$.
In addition all the vertices will be chosen with definite 
(anti-) self-duality properties.
The computation and notations follow closely \cite{agnt}.    

The vertex operators for the gauge field strengths (\ref{flr})
are given by
\bea
V_L(p,\xi)&=&\int d^2 z \,\xi_{\mu i}\,(\partial X^\mu-i p \chi \chi^\mu)\,
(\bar{\partial} X^i-i p\tilde{\chi}
\tilde{\bar{\chi}}^i)\,e^{i p X}\nonumber\\
V_R(p,\xi)&=&\int d^2 z \xi_{\mu i}\,(\partial X^i-i p \chi \chi^i)\,
(\bar{\partial} X^\mu-i p \tilde{\chi} \tilde{\chi}^\mu)\,
e^{i p X}
\label{verticeii}
\eea
Notice that each vertex carries at least one power of space-time
momentum $p_\mu$, and therefore to the order we 
are interested in we can keep only linear
terms in $p_{\mu}$.
                                                                              
A representative of such couplings
in each of the three models is indicated in the table below, 
where we also indicate how the various charges are 
mapped to each others:

$$
\begin{CD}
II_F @>T_{15}ST_{2345}S>> I_F @>T_{15}ST_{2345}S>> III_F\\
(-)^{F_L}\sigma_6 &&  (-)^{F_L}I_{4}\sigma_6 
&& I_{4}\sigma_6\\
F_1 && P_1 && NS_{12345}\\
NS_{12345}&& F_1 && P_1\\
P_1 && NS_{12345} && F_1\\
({\cal F}^+_R)^2({\cal F}^-_R)^2({\cal F}^+_L)^{2k} 
&&  
({\cal F}^+_{{R}})^2({\cal F}^+_{{L}})^2
({\cal F}^+_L)^{2k}  
&& ({\cal F}^-_{L})^2({\cal F}^-_{{R}})^2
({\cal F}^+_L)^{2k}\\
\end{CD}
$$
The field strengths 
${\cal F}^{\pm}_{L,R}$ are defined in (\ref{flr}) and $\sigma_6$
is a shift of order 2 along the 6th direction.

The couplings in the table are special in the sense that they receive
in each case contributions only from right moving ground states
(BPS saturated states). 
This can be seen 
by noticing that the insertions exactly soak the number of fermionic
zero modes in the right-moving part of the string amplitudes. 
This will be implicit in most of our discussion. 

In the model $II_F$ this corresponds
to the case where the eight right moving fermionic zero
modes (once the sums over spin structure have been performed) are
soaked up by exactly four insertions $({\cal F}_R^+ {\cal F}_R^-)^2$
of right moving gauge fields. Similarly, in the models $I_F$ and $III_F$
two right-moving insertions of  $({\cal F}_R^+)^2$ 
and $({\cal F}_R^-)^2$ respectively
are needed in order to get a non-trivial result.
Vertex operators for self-dual components can therefore be 
effectively replaced by:  
\bea
\vec{V}_{L}^{+}(p_1)&=&i p_1 \vec{P}_L \tau_2\,\int d^2 \sigma 
(Z^1\bar{\partial} Z^2-\tilde{\chi}^1 \tilde{\chi}^2)+...\nonumber\\
\vec{V}_{L}^{+}(\bar{p}_2)&=&i \bar{p}_2 \vec{P}_L\tau_2\,\int d^2 \sigma 
(\bar{Z}^2\bar{\partial}\bar{Z}^1-\tilde{\bar{\chi}}^2 
\tilde{\bar{\chi}}^1)+...\nonumber\\
\vec{V}_{R}^{+}(p_1)&=&i p_1 \vec{P}_R \tau_2\,\int d^2 
\sigma (Z^1\partial Z^2-\chi^1 \chi^2)+... \nonumber\\
\vec{V}_{R}^{+}(\bar{p}_2)&=&i \bar{p}_2 \vec{P}_R\tau_2\,
\int d^2 \sigma 
(\bar{Z}^2\partial \bar{Z}^1-
\bar{\chi}^2 \bar{\chi}^1)+...
\label{veff}
\eea
where we have grouped the components $P^i_L=\partial X^i$, 
$P^i_R=\bar{\partial} X^i$ with $i=1,2$ into a two-dimensional
vector and $\partial=
{1\over \tau_2}(\partial_{\sigma_2}-\bar{\tau}\partial_{\sigma_1})$.  
Similar expressions are given for anti-self-dual components, replacing
$Z^2,\chi^2,\tilde{\chi}^2$ by their complex conjugates. 

From the expressions for the effective vertex operators
in (\ref{veff}), we see that their insertion in the
correlator (\ref{ak}) amounts to inserting factors of $\vec{P}_{L,R}$.

Since the vertices are quadratic in the quantum fluctuations
one can exponentiate them into a generating function 
\be
{\cal G}_{\vec{a},\vec{b}}(v,w)=
\big\langle e^{-S_0-\vec{v}\cdot \vec{V}_L-\vec{w}\cdot \vec{V}_R }
\big\rangle= 
\int {d^2 \tau\over \tau_2^3}\,\sum_{g,h}\, q^n\, C{g\brack h} (n,{\bf l})
\,\Gamma_{d,d}{g\brack h}(\ell\cdot v,\ell\cdot w)  
\label{g123}
\ee
with $S_0$ the free string action, 
$q=e^{2\pi i\tau}$ ($\tau$ 
the genus-one worldsheet modulus) and $\vec{v}_{\pm},\vec{w}_{\pm}$ are 
sources for the eight $U(1)$ gauge fields. The 
scalar product is defined as usual by $\vec{a}\, \vec{b}=a_1b_1+a_2b_2$,
while the dotted product reads $a\cdot b=a_+b_- +a_-b_+$.
In the right-hand side we have introduced the notation 
$v_{\pm}\equiv \vec{v}_{\pm}\vec{P}_{L}$ with a similar
definition for $w_{\pm}$ with $\vec{P}_{L}$ replaced by $\vec{P}_{R}$
and $\vec{v}$ by $\vec{w}$. 
Vertex insertions are defined by   
$\vec{v}_{\pm},\vec{w}_{\pm} $-derivatives of (\ref{g123}). 

Finally, we denote as before by $C{g\brack h} (n,{\bf l})$
with  
${\bf l}\equiv(\ell,\tilde{\ell},
\ell_*,\tilde{\ell}_*)=(\ell^+,\ell^-,\ell_*^+,\ell^-_*)$
the coefficients in the expansion of the partition function
which includes Wilson lines:
\bea
G{g \brack h}(q,{\bf y})
&=&{\rm Tr}_{g-tw}^\prime \left[\Theta^h\, q^{L_0-c/24}\,
\bar{q}^{\bar{L}_0-c/24}\,
y^{2 J^3_{0}}\,\tilde{y}^{2 \tilde{J}^3_{0}}\,
y_*^{2 \bar{J}^3_{0}}\,\tilde{y}_*^{2 \tilde{\bar{J}}^3_{0}}\right]
\nonumber\\
&=& C{g\brack h} (n,{\bf l}) \,
q^n\, y^{\ell}\tilde{y}^{ \tilde{\ell}}\,
y_*^{\ell_{*}}\,\tilde{y}_*^{\tilde{\ell}_{*}}
\label{helicity}
\eea
$\Theta$ is the $Z_2$ orbifold
generator, $y=e^{\pi i v_-}$, 
$\tilde{y}=e^{\pi i v_+}$,
and similarly  $y_*,\tilde{y}_*$ are obtained by replacing
$\vec{v}_\pm$ by $\vec{w}_\pm$.
$L_0$, $\bar{L}_0$ are the Virasoro generators and
$J^3_{0}$'s are four $U(1)$
generators to which the
corresponding gauge fields couple. Therefore we see that
(\ref{helicity}) has the structure of a helicity supertrace.
The four possible twists
along the $\sigma$ and $\tau$ directions 
will be denoted by ${g\brack h}$ with $g,h=0,{1\over 2}$. 
The primes denote the omission of the bosonic
zero mode contributions which have been displayed explicitly
in (\ref{g123}). $\Gamma_{d,d}$ is a $\Gamma_{2,2}\times\Gamma_{4,4}$ lattice 
sum for the completely untwisted sectors in the models $I_F,III_F$
and all sectors in the model $II_F$, while reduced to a 
$\Gamma_{2,2}$ lattice for all non-trivial twists in models
$I_F,III_F$.
Since we are interested only in the $(T,U)$ moduli dependence of the
first torus we will always work in the orbits where neither momentum 
nor winding
modes are excited in the $\Gamma_{4,4}$-lattice.
The effects of introducing half-shifts in $\Gamma_{d,d}$ lattices have been 
extensively studied in \cite{ko}. 
The perturbed lattice sum can be written as the sum
\bea
\Gamma_{2,2}{g\brack h}(\ell\cdot\vec{v},\ell_*\cdot\vec{w})&=&
\frac{T_2}{\tau_2}\,
 \sum_{M}\eta{g\brack h}
e^{- \frac{\pi T_2 }{ \tau_2 U_2 }
\big| (1\; U)M  \big( {\tau \atop -1} \big) \big| ^2}\nonumber\\
&& 
\times e^{2\pi i\left[ T {\rm det}M+(\ell\cdot v^1\; , 
\ell\cdot v^2)
M  \big( {\tau \atop -1} \big) 
-(\ell_*\cdot w^1\; , \ell_*\cdot w^2)M  
\big( {\bar{\tau} \atop -1} \big)\right]}
\label{latpert}
\eea
over worldsheet instantons
\bea
\pmatrix{X^1\cr X^6}=
 M \pmatrix{\sigma^1\cr \sigma^2}\equiv
\pmatrix{m_1&n_1 \cr
m_2&n_2}\pmatrix{\sigma^1 \cr \sigma^2}\quad 
\vec{m}\in {\bf Z}+\vec{b} g,\,\vec{n}\in {\bf Z}+\vec{b} h,\,
\label{M}
\eea
where now the entries in $M$ are integer or half-integer depending
on the winding(momentum) shift vector 
$\vec{a}$ ($\vec{b}$).
Finally the lattice sum is weighted by the $\eta {g\brack h}$ phase 
\be
\eta {g\brack h}=e^{-4\pi i\vec{a}\vec{b}\,gh-2\pi i
\vec{a}\,(h \vec{m}-g\vec{n})}     
\ee
We will consider only shifts involving
either a pure momentum or pure winding $\vec{a}\,\vec{b}=0$.

Evaluating the partition functions (\ref{helicity}) in the three
orbifold models described above one is left with (after performing 
the spin structure sums): 
\bea
G_{II_F}{g\brack h}(q,{\bf y})
&=&
\frac{\vartheta{g\brack h}^2(y)
\vartheta{g\brack h}^2(\tilde{y})}
{\hat{\vartheta}_1(y \tilde{y})
\hat{\vartheta}_1(y \tilde{y}^{-1})\eta^6}\, 
(y_*^{1\over 2}-y_*^{-{1\over 2}})^2 (\tilde{y}_*^{1\over 2}-
\tilde{y}_*^{-{1\over 2}})^2
\nonumber\\
{G}_{I_F}{g\brack h}(q,{\bf y})&=&\frac{\vartheta{g\brack h}^2(y)
\vartheta^2_1(\tilde{y})}
{\hat{\vartheta}_1(y \tilde{y})
\hat{\vartheta}_1(y \tilde{y}^{-1})\hat{\vartheta}{g\brack h}^2(0)}\, 
(y_*^{1\over 2}-y_*^{-{1\over 2}})^2 \nonumber\\
{G}_{III_F}{g\brack h}(q,{\bf y})&=&\frac{\vartheta^2_1(y)
\vartheta{g\brack h}^2(\tilde{y})}
{\hat{\vartheta}_1(y \tilde{y})
\hat{\vartheta}_1(y \tilde{y}^{-1})\hat{\vartheta}{g\brack h}^2(0)}\, 
(\tilde{y}_*^{1\over 2}-
\tilde{y}_*^{-{1\over 2}})^2 
\label{g123t}
\eea
The hat on the $\vartheta$-functions in the denominators
denotes as before the omission of their zero mode parts, i.e. 
$\hat{\vartheta}_1(v)\equiv {1\over v}\vartheta_1(v)$, 
$\hat{\vartheta}_2(v)\equiv {1\over 2}\vartheta_2(v)$ and 
$\hat{\vartheta}_{3,4}(v)\equiv \vartheta_{3,4}(v)$.  

Notice the modular invariance of ${\cal G}(v,w)$
under the SL(2,Z) transformations
\be
\tau\rightarrow {p\tau+q\over r\tau+s}\quad
v\rightarrow {v\over r\bar{\tau}+s}
\quad w\rightarrow {w\over r\tau+s}
 \quad M\rightarrow M \pmatrix{s & q \cr r & p}
\ee

The modular integral (\ref{g123}) can then be computed 
following the standard trick \cite{dkl}, 
that consists in trading the
sum over the $M_{\epsilon}$ matrices in (\ref{latpert}) by sums over 
$SL(2,Z)$ representatives integrated in unfolded domains. 
We will concentrate here on the contributions
of non-degenerated orbits (${\rm det}\, M\neq 0$) for which
representatives can be chosen to be:
\be
M  = \left( \matrix{m_1  & n_1\cr 0  & n_2 \cr}\right)
\label{Mnondeg}
\ee  
where $m_1\in{\bf Z}+b_1 g$ and $n_1\in{\bf Z}+ b_1 h$. 
The integral (\ref{g123}) is then unfolded to the whole upper half plane.

The modular integral (\ref{g123}) is evaluated in the Appendix A.
We keep only its leading order in the expansion 
around $T_2\rightarrow \infty$. In the dual picture, this corresponds  
to the classical contribution of the D-instanton background.
Higher orders can be interpreted  as quantum
fluctuations around the instanton background as in \cite{ghmn}, but
the analysis of these terms is beyond the scope of this work.

The final result, formula (\ref{final}) of the Appendix C, 
after performing the $m_1$ and $n_2$ sums 
can be written as:
\be
I(a,b)={\rm ln} \, {\cal Z}(a,b)\bar{{\cal Z}}(a,b)
\ee
with
\bea
{\cal Z}(1,0)=\prod_{\delta=0}
(1-p^{m_1}q^{k} \hat{y}^{\ell}
\tilde{\hat{y}}^{\tilde{\ell}} \hat{y}_*^{\ell_*}
\tilde{\hat{y}}_*^{\tilde{\ell}_*} )
^{C_+\{{k\over 2}\}(k m_1,{\bf l})}&&\nonumber\\
\times\prod_{\delta={1\over 2}}
(1-p^{m_1}q^{k} \hat{y}^{\ell}
\tilde{\hat{y}}^{\tilde{\ell}} \hat{y}_*^{\ell_*}
\tilde{\hat{y}}_*^{\tilde{\ell}_*} )
^{C_-\{{k\over 2}\}(k m_1,{\bf l})}
&&\nonumber\\
{\cal Z}(0,1)=\prod_{\delta=0}
(1-p^{2 m_1} q^{{k\over 2}} \hat{y}^{\ell}
\tilde{\hat{y}}^{\tilde{\ell}} \hat{y}_*^{\ell_*}
\tilde{\hat{y}}_*^{\tilde{\ell}_*} )
^{C_+\{{m_1\over 2}\}(k m_1,{\bf l})}&&\nonumber\\
\times\prod_{\delta={1\over 2}}
(1-p^{2m_1} q^{k\over 2} \hat{y}^{\ell}
\tilde{\hat{y}}^{\tilde{\ell}} \hat{y}_*^{\ell_*}
\tilde{\hat{y}}_*^{\tilde{\ell}_*} )
^{C_-\{{m_1\over 2}\}(k m_1,{\bf l})}
\label{genusf}
\eea
The various quantities entering in (\ref{genusf}) have the following 
meaning: $p=e^{2\pi i T}$ and $q=e^{2\pi i U}$, 
$\hat{y}=e^{\pi i \hat{v}}$ and $\hat{y}_*=e^{\pi i \hat{w}}$
are the induced sources, with  
$\hat{v}=v_1 U-v_2$ and $\hat{w}={1\over U_2} (w_1 \bar{U}-w_2)$  
and similar definitions for $\tilde{\hat{y}}$ and
$\tilde{\hat{y}}_*$. 
Finally $\bar{\cal Z}$ is simply the complex conjugate 
of ${\cal Z}$ and represents the anti-instanton contributions.

Notice that ${\cal Z}(1,0)$ and ${\cal Z}(0,1)$ in (\ref{genusf})
are mapped into each other under the simultaneous exchange of the
momentum($b_1$)-winding($a_1$) shifts and $k$-$m_1$ modes, as required
by T-duality. 

This formula is rather more general than what we really need. 
Still, depending on the model, a certain number  of ${y}$-derivatives 
should be taken and then the right moving source $y_*$ should be set to
zero ($y_*=\tilde{y}_*=1$). Notice however that already at this stage 
one can recognize in ${\cal Z}(0,1)$ the symmetric product formula (\ref{11})
for the longitudinal shift elliptic genus. 
This is enough to ensure, following \cite{bachas, bgmn, ghmn}, the
agreement between the D-instanton corrections associated to
the states counted by (\ref{11}), provided     
that the orbifold CFT describing the D1(D5)-KK 
system is constructed
out of the symmetric product of $N$ copies 
of the fundamental theory in the twisted sector (basic unit of winding).

Coming back to our formula, 
after acting in (\ref{genusf}) with the appropriate 
number\footnote{The minimal one in order to get a non-trivial result} 
of $w$-derivatives (see \cite{mms}
for similar manipulations) one is left with:
\bea
\hat{{\cal Z}}(1,0)&=&
{1\over 2}\sum \,  
(-)^{m_1 h}\, \hat{C}{g\brack h}(k m_1,\ell,\tilde{\ell}) 
\,p^{m_1 s}\, q^{k s}\, y^{s \ell}\, 
\tilde{y}^{s \tilde{\ell}}\nonumber\\
\hat{{\cal Z}}(0,1)&=&
{1\over 2}\sum \,(-)^{k h}\, 
\hat{C}{g\brack h}(k m_1,\ell,\tilde{\ell})\, p^{m_1 s}\, q^{k s}\,
 y^{s \ell} \,\tilde{y}^{s \tilde{\ell}}
\label{zmn}
\eea
where 
\be
\hat{{\cal Z}}(a,b)\equiv {1\over m! n!}
{\partial^m\over \partial y_*^m}{\partial^n\over \partial\tilde{y}_*^n}
{\cal Z}(a,b)|_{y_*=\tilde{y}_*=1}
\ee
with $m=2,n=0$ for the model $I_F$, $m=n=2$
for the model $II_F$ and $m=0,n=2$ for the model $III_F$. The sum 
run over $\ell,\tilde{\ell}$ integers, 
$g,h=0,{1\over 2}$,
$m_1\in {\bf Z}+b_1 g$ and $k\in {\bf Z}+a_1 g$. 
Finally, the coefficients $\hat{C}{g\brack h}(\Delta,\ell,\tilde{\ell})$ 
are similarly defined in terms of the expansion coefficients of
(\ref{g123t}) by  
\be
\hat{C}{g\brack h}(\Delta,\ell,\tilde{\ell})\equiv
\sum_{\ell_*,\tilde{\ell}_*}\, (-)^{k h}\, \ell_*^m\, \tilde{\ell}_*^n  
\,C{g\brack h}(\Delta,\ell,\tilde{\ell},\ell_*,\tilde{\ell}_*).
\ee
and correspond to the expansion coefficients of the chiral supertraces
appearing in (\ref{g123t}).

\section{Conclusions and open problems}

The main goal of this paper has been the formulation of
CFT descriptions of the moduli space of 
D1/D5-brane systems in a class of 
models with 16 supercharges. These models were obtained by 
orbifolding/orientifolding type IIB theory accompanied by
a $Z_2$ shift. The presence of the $Z_2$ shift allowed us 
to apply the adiabatic principle in order to obtain the corresponding
CFTs.
Our proposed CFTs involved
symmetric products of $R^4$ and $T^4$ factors with 
additional $Z_2$ actions,
whose precise form depends on the background in consideration.
For backgrounds involving $\Omega$ projection, the CFTs
turn out to be $(4,0)$.

We have worked out elliptic genus formulae for these
modified symmetric products and shown that the resulting
multiplicities for D1/D5 bound states were in agreeement 
with those of winding/momentum states in U-dually related
theories.

There remain however several open problems,
which we have already anticipated in the introduction 
and discussed in Section 5. 

Probably the most challenging one 
concerns the issue of 3-charge
systems. As stressed above, our CFTs predict multiplicities
which agree with U-duality in the 2-charge cases, i.e.
in the CFTs of D1(D5)-KK systems or pure D1/D5 
bound states, i.e. $\Delta={\bar\Delta}=0$.  
In general, problems arise when
we excite momentum in the D1/D5-brane system, that is 
we let  $\Delta\neq 0$. This 
corresponds to exciting states
which preserve $1/4$ of the 16 bulk supercharges 
(in the $(4,0)$ case the
momentum is excited in the right-moving, non-supersymmetric
sector). As we have noted  in section 5, 
U-duality in this case puts constraints which generically are not
satisfied by the proposed CFTs. To put it even more
dramatically, the predictions of U-duality apparently
do not admit any CFT interpretation, as they clash with
modular invariance.
Since this problem is generic to all three charge
systems in theories with sixteen supercharges studied so far,
including the D1/D5 brane system in type IIB theory on $K3$, let us
summarize the assumptions involved. 

The usual type IIB
theory on $K3$ as well as the model $II$ studied in this paper are (4,4)
superconformal field theories. In the $K3$ case there are good
reasons to believe that the CFT is a deformation of the (4,4)
symmetric product CFT times the center of mass CFT. Since model $II$ is
closely related to $K3$, it is also reasonable to assume a symmetric
product (4,4) CFT. The elliptic genus (after  taking 2-derivatives
with respect to $\tilde{y}$ and setting it to 1) is then of the form
\be
Z(p.q,y)= Z_{cm}(q,y) Z_{sym}(p,q,y)~~~~~~~
Z_{sym}(p,q,y)={\hat Z}_F(p,y)\hat{Z}_{sym}(p,q,y)
\ee
where the subscripts $cm$ and $sym$ denote center of mass and the
internal theory (which in this case we are assuming is a symmetric
product CFT)
respectively and in the second equation we have separated out the $q^0$
term in ${\hat Z}_F$ and the remaining terms in $\hat{Z}_{sym}$. 
An inspection of the
second equation in (\ref{11}) relevant for the transverse shift case,
shows that $\hat{Z}_{sym}$ is symmetric under the exchange of $p$ and $q$.
U-duality which exchanges D1 and KK modes, will exchange $p$ and $q$,
with the result that $Z_F$ now becomes the center of mass contribution
$Z'_{cm}$ and the internal contribution $Z'_{in}$ (which we don't
assume to be necessarily a symmetric product CFT) is the
product $Z_{cm}(p,y)\hat{Z}_{sym}(q,p,y)$. 
Expanding $Z'_{in}$ in powers of $p$, at
each order $p^n$ the coefficient $Z'_n(q,y)$ must describe the internal (4,0)
CFT of the system of 1 D5 and $n$ D1 branes in the U-dual theory.  
In particular for $n=1$ we find
\be
Z'_1(q,y) = Z_1(q,y) - Z_1(0,y) + f_1(y)
\ee
where $f_1$ is coefficient of $q^1$ in the $q$-expansion of
$Z_{cm}(q,y)$ and $Z_1$ is the elliptic genus for the single copy of
the (4,4) internal theory.
Now $Z_1$ is a modular form in $q$ and $y$ under a suitable subgroup
of $SL(2,Z)$ acting in the usual way. Thus, unless $f_1(y)= Z_1(0,y)$,
the left hand side $Z'_1(q,y)$ will not be a modular form. On the
other hand, $f_1(y)$ cannot be equal to $Z_1(0,y)$, since at $y=1$ the
former vanishes as it describes the bound states of D1/D5 system
which forms 1/4 BPS state, while the latter is non-zero since it
describes 1/2 BPS bound state. One might wonder that since in the
U-dual theory we have only (4,0) supersymmetry, the $SU(2)$ to which
$y$ couples is broken. However, this is part of the $SO(4)_E$ which is
the little group of the system under consideration and therefore it
should have a well defined action on the states. If we assume that at
the conformal invariant fixed point this global symmetry should be
promoted to Kac-Moody algebra in the internal theory 
(actually we need only the $U(1)_y$
current algebra to which $y$ couples), then $Z'_1(q,y)$ must be a
modular form. Thus we conclude that (4,4) symmetric product theory,
U-duality and the existence of $U(1)_y$ current algebra in the internal
part of the U-dual (4,0) CFT cannot be satisfied simultaneously.
 
One may try to argue that the problem is related to the fact that
the 1/4-BPS states in theories with 16 supercharges generically are
not stable throughout the whole moduli space. As a result the
elliptic genus might jump. 
For example, it is
believed \cite{N4} that in ${\cal N}=4$ 
Yang-Mills theory in 4 dimensions
$1/4$ BPS dyonic states do indeed decay after crossing codimension 1
regions
of marginal stability in the moduli space. In other words, single
particle BPS states become multiparticle states. 
One can see from the mass formula for 1/4-BPS states that the region
of marginal stability in 4-dimensional string theories is of real
codimension 1 while in 5-dimensional theory it is the entire moduli space. 
This is indeed the case for  
the 5-dimensional type I' theory (which is U-dual to IIB on
$K3$) as well as 
model $III$ where the D1/D5 system is 1/4-BPS \footnote{Actually
model $III$ with transverse shift is a 4-dimensional model and the
subspace at which the system becomes threshold is real codimension one
in the moduli space of this 4-dimensional theory. However the modulus 
which takes one away from the singular subspace
is
a certain combination of the metric $G_{16}$ and the RR 2-form
$C_{16}$
which would break the Lorentz invariance in the world sheet along 01 
directions.} .
However the D1/D5 system in the model $II$ (or IIB on $K3$) is 1/2
BPS. This
system is at threshold in a codimension 4 subspace and as a result
the
corresponding (4,4) CFT is non-singular. This would suggest that the
elliptic genus for this model should be constant throughout the
moduli-space, which in turn would indicate
that the 3-charge system is stable everywhere.

The symmetric product CFTs which arise  by the adiabatic
argument
presented in section 3, is presumably valid at 
$\chi=C_{2345}/Q_1=1/2$.
However, unlike in model $II$,  the $\chi$ and $C_{(4)}$ fields are
projected out in model $III$,
and therefore, the $\chi=1/2$ point 
cannot be connected to the point $\chi=0$. In fact, as
discussed at the end of section 2, 
the models obtained by $\Omega$ projection at $\chi=1/2$ may
be quite different from the one at $\chi=0$.  This might be a possible
explanation for why the elliptic genus of the symmetric product theory
for model $II$ after $p$, $q$
exchange,
which by U-duality should describe the multiplicities of the 3 charge
system of model $III$ at $\chi= C_{2345}=0$, differs from that of the
symmetric product theory for model $III$ which is valid at nonzero $\chi$
and RR 4-form field. Conversely,$\chi$ and RR 4-form $C_{2345}$ of model
$III$, under U-duality are mapped to NS 2-form $B_{15}$ and the RR
4-form $C_{1234}$ in model $II$. Note that these fields are projected
out in model $II$ but their discrete $Z_2$ fluxes are allowed.
Therefore under the U-duality the
symmetric product CFT of model $III$ should be compared with the 3
charge system in model $II$ in the background of the latter fields, which
however breaks the Lorentz invariance of the world sheet along
01 directions.

The near horizon geometries of the D1/D5 systems describe $AdS_3\times
S^3$ times a 4-dimensional internal space. 
One can use AdS/CFT correspondence
to compare the results from the bulk with that of the CFTs studied
here. By introducing the concept of {\it degree},
\cite{DB2},
this comparison can be extended also to non-chiral primaries.
In \cite{ghmn1}, we show that for model $II$, the CFT results are in
complete agreement with the AdS/CFT predictions. On the other hand for
model
$III$, supergravity result is in contradiction with the (4,0) CFT for
the excited states. In fact, it turns out that supergravity elliptic genus is
consistent with the U-dual model, namely it reproduces the elliptic
genus of model $II$ with $p$ and $q$ exchanged (of course in the regime
of validity). This result seems to support the remarks in the
previous paragraph, according to which the elliptic genus for model
$III$ at $\chi=0$ should be given by that of model $II$ 
(after $p$, $q$ exchange). It is however hard to see how the gravity
analysis in model $III$ would be effected by the non-trivial
backgrounds of $\chi$ and the RR 4-form field. 
Therefore, a proper understanding of the (4,0) models remains 
still an important open question.

\vskip 0.5in
{\bf Acknowledgements}

We acknowledge discussions with J. de Boer, E. Kiritsis, J. Maldacena, 
B. Pioline, E Verlinde and expecially with
M. Bianchi and G. Thompson at various stages of this work.
This project is supported in part by EEC under TMR contracts
ERBFMRX-CT96-0090, HPRN-CT-2000-00148 ,
HPRN-CT-2000-00122 and the INTAS project
991590. The work of J.F.M. is supported by the INFN
section of University of Rome ``Tor Vergata''.

\vskip 0.5in

\section{Appendix A: $SO(n_1)\times SO(n_5)$ D1/D5 gauge theory}
\renewcommand{\theequation}{A.\arabic{equation}}
\setcounter{equation}{0}

In this appendix we determine the spectrum of 
open string states living  
on intersections of D1/D5-branes,  
in ``type I'' like backgrounds,
where the orientifold group action is accompanied
by a shift longitudinal to the world-volume system. 
Our aim is to show that, unlike for the 
more familiar type I cousins,
where consistency of the underlying open string theory 
requires that $\Omega$ projection acts with a relative
sign between the D1 and D5 gauge groups \cite{gp}, 
in the presence of a longitudinal shift $SO(n_1)\times SO(n_5)$ Chan-Paton 
assignements are allowed. We adopt the open string 
descendant techniques systematized in \cite{torvergata}
\footnote{For a quick review of these techniques 
and applications close
to the ones presented here, see sections 2 of \cite{bmAT} and 3 of 
\cite{bmRG}. A systematic study of type I string 
vacua involving D-branes in the
presence of shifts can be found in \cite{sagnotti}.}.   
Being interested in open string theories describing excitations 
of D-brane bound states rather than vacuum configurations we
relax (and generically violate) tadpole cancellation conditions.  
We will discuss the case of model $I$, with orientifold group
generated by $\Omega\sigma_{p_1}$, but minor modifications 
are required to describe the $T$-dual model $III$ associated to
$\Omega I_4\sigma_{p_1}$.  

These vacuum configurations are often termed
as ``type I theory without open strings'', 
since the Klein bottle tadpole is removed by the presence of the shift
and therefore the inclusion of D9(D5)-branes with their 
corresponding open string
excitations are no longer needed \cite{massimo}.
Although we are mainly interested in the study of pure D1/D5 systems,
with a little more effort, we can (and we will) include also 
$n_9$ D9-branes
in our analysis. Besides aesthetical reasons, the inclusion of D9-branes
will help the comparison with the more familiar type I results.   

We start by describing the D1-D5-D9 system in the presence of the
standard type I orientifold (O9)-plane. We orient
$n_1$ D1- and $n_5$ D5-branes along $(01)$ and $(012345)$ planes
respectively. For an homogeneous notation 
it will be convenient to start by
wrapping the whole system on 
a $S^1\times T^4\times \tilde{T}^4$ torus, with directions 
$(1)\times (2345)\times (6789)$, and only at the end take the 
volume of $\tilde{T}^4$ to infinity. 
The annulus, Moebius strip and Klein bottle amplitudes 
associated to such brane configuration can be
written as $\int {d t\over t^{3\over 2}}\times$ 
\bea
{\cal K}&=&{1 \over 2}\,\rho_{00}(2it)P_1(t)P_4(t) \tilde{P}_4(t)\nonumber\\ 
{\cal A}&=&{1 \over 2}\,\left[\rho_{00}\left({it\over 2}\right)\left(
n_9^2\,P_4(t) \tilde{P}_4(t)+n_5^2\,P_4(t) \tilde{W}_4(t)
+n_1^2\,W_4(t) \tilde{W}_4(t)\right)\right.\nonumber\\
&&\left.
+2 n_5 n_1 \rho_{A0}\left({it\over 2}\right)\tilde{W}_4(t)
+2 n_5 n_9 \rho_{B0}\left({it\over 2}\right)\times\right.\nonumber\\
&&\left. 
P_4(t) 2 n_9 n_1 \rho_{C0}\left({it\over 2}\right)\right]P_1(t)\label{akm}\\
{\cal M}&=&{1 \over 2}\,\left[-n_9\,
\rho_{00}\left({it\over 2}+{1\over 2}\right)
P_4(t)\tilde{P}_4(t)
+n_5\,\rho_{0B}\left({it\over 2}+{1\over 2}\right)P_4(t)
\right.\nonumber\\
&&\left. -n_1\,\rho_{0C}\left({it\over 2}+{1\over 2}\right)\right]P_1(t)
\nonumber
\eea  
where $A$, $B$ and $C$ in $\rho_{gh}$ refers to $h$ projection
of the $g$-twisted chiral traces with twists oriented 
along the planes $(2345)$,
$(6789)$ and $(23456789)$ respectively. After 
performing the spin
structure sums these traces can be written as
\bea
\rho_{gh}&=&{\vartheta_1^2\over \eta^6}{\vartheta^2
{{1\over 2}+g\brack {1\over 2}+h}\over \hat{\vartheta}^2
{{1\over 2}+g\brack {1\over 2}+h}}~~~~g,h={1\over 2}~~{\rm for}~~
g,h=A,B\nonumber\\ 
\rho_{gh}&=&{\vartheta^4
{{1\over 2}+g\brack {1\over 2}+h}\over \hat{\vartheta}^4
{{1\over 2}+g\brack {1\over 2}+h}}~~~~~~~~~g,h={1\over 2}~~{\rm for}~~
g,h=C
\label{rho}
\eea
Finally momentum and winding lattice sums are given by
\be
P_{d}(t)=\sum_{m_i\in {\bf Z}^d}\, 
e^{-\pi t \alpha^\prime {m_i^2\over R_i^2}}
\quad\quad
W_{d}(t)=\sum_{n_i\in {\bf Z}^d}\, 
e^{-\pi t n_j^2 {R_j^2 \over \alpha^\prime}}
\label{pw}
\ee 
The basic requirement that these string amplitudes (\ref{akm})   
should satisfy, is that, after the exchange of $\sigma$ and $\tau$ 
directions, they must admit a sensible interpretation in terms 
of closed string exchanges between boundaries (D-branes)
and crosscaps (O9-planes). More precisely, the sum
of closed string amplitudes should reconstruct the whole square 
\be
\langle B|\, e^{-l H}\,|B \rangle
\label{bb}
\ee
with  
\be
|B \rangle=|O9\rangle+n_9\, |D9\rangle+n_5\, |D5\rangle +
n_1\, |D1\rangle .
\ee
and $|O9\rangle$, $|D9\rangle$, $|D5\rangle$ and $|D1\rangle$ representing
the boundary state for corresponding brane objects.
 
Rewriting (\ref{akm}) in terms of the closed string
variables $\ell_K={1\over 2t}$, $\ell_A={2\over t}$,
$\ell_M={1\over 2t}$ one is left (at the origin of the 
$T^4\times \tilde{T}^4$ lattice sum) with $\int d\ell\times$
\bea
\widetilde{\cal K}_0 &=&{2^5\over 2}\left[\chi_O+\chi_V+\chi_S+\chi_C \right] 
(i\ell)\, v_1 v_4 \tilde{v}_4 \,W_1^{\rm even}({\ell\over 2})\nonumber\\
\widetilde{\cal A}_0 &=&{2^{-5}\over 2}\left[ 
\chi_O I_O^2+\chi_V I_V^2 +\chi_S I_S^2+\chi_C I_C^2 \right] 
(i\ell)\, W_1({\ell\over 2})\nonumber\\ 
\widetilde{\cal M}_0 &=&-{2\over 2}\left[
\chi_O I_O+\chi_V I_V +\chi_S I_S+\chi_C I_C \right] 
(i\ell)\,\sqrt{v_1 v_4 \tilde{v}_4}\, W_1^{\rm even}({\ell\over 2})
\label{kamt}
\eea
$W_1^{\rm even}$ is defined like (\ref{pw}), with $n$ restricted to be 
even. 
We have introduce the linear combinations of $\rho_{gh}$
traces
\be
\pmatrix{\chi_{0}\cr \chi_{V}\cr \chi_{S}\cr \chi_{C}}
={1\over 4}\pmatrix{+& +& +& +\cr +& +& -& -\cr +& -& -& +\cr +& -& +& -\cr} 
\pmatrix{\rho_{00}\cr \rho_{0A}\cr 
\rho_{0B}\cr \rho_{0C}}
\label{sso8}
\ee
and the Chan-Paton dependent combinations
\be
\pmatrix{I_{0}\cr I_{V}\cr I_{S}\cr I_{C}}
=\pmatrix{+& +& +\cr +& -& -\cr +& -& +\cr +& +&-\cr} 
\pmatrix{n_9\sqrt{v_1 v_4 \tilde{v}_4}
\cr n_5 \sqrt{v_1 v_4\over \tilde{v}_4}     \cr 
n_1\sqrt{v_1 \over v_4 \tilde{v}_4} }.
\label{io8}
\ee
with $v_1$, $v_4$ and $\tilde{v}_4$ 
the volumes of $S^1$, $T^4$ and  $\tilde{T}^4$ respectively.
The relevant modular transformations connecting the two 
expressions (\ref{akm}) and (\ref{kamt}) can be easily read from 
(\ref{rho}) to be: 
\bea
&&\rho_{00}(-{1\over i\alpha \ell})=(\alpha \ell)^{-4}\rho_{00}(i\alpha\ell)
\quad \rho_{00}({i\over 2 t}+{1\over 2})=t^{-4}\rho_{00}
({it \over 2 }+{1\over 2})\nonumber\\
&&\rho_{0 h}(-{1\over i\alpha \ell})=4 (\alpha \ell)^{-2}\rho_{h0} 
(i\alpha\ell)\quad 
\rho_{0h}({i\over 2 t}+{1\over 2})=-t^{-2}\rho_{0h}
({it \over 2 }+{1\over 2})
\nonumber\\
&&\rho_{0 C}(-{1\over i\alpha \ell})=
16\rho_{C0}(i\alpha\ell)\quad
\rho_{0C}({i\over 2 t}+{1\over 2})=\rho_{0C}
({it \over 2 }+{1\over 2})\nonumber\\
&&~~~~~~~~~~~~~~~~~~~~~~~~~P_d({1\over \alpha\ell})=v_d 
(\alpha \ell)^{d\over 2} W_d(\alpha\ell)
\eea
with $h=A,B$ and $\alpha$ a factors of $2$ depending on the 
kind of one-loop diagram.

Rewritten in the basis (\ref{sso8}), one can easily recognize
in $\widetilde{\cal K}_0+
\widetilde{\cal A}_0+\widetilde{\cal M}_0$ given by (\ref{kamt})
as the different terms in  
the square (\ref{bb}). Notice that, unlike tadpole
cancellation conditions, the requirement
that the whole amplitude reconstructs a square, 
is a restriction on the structure of
the entire tower of massive closed string amplitudes. 
Indeed, this requirement together with the choice 
of SO gauge groups for D9-branes is sufficient to fix 
completely the Moebius strip string amplitudes, 
once the Klein bottle and
annulus amplitudes are given \cite{torvergata}. In particular,
the relative signs between the $\Omega$ projections
on D5 and D1, D9 Chan-Paton factors are crucial in order to
reproduce the square.     

Let us consider the same  system in the ``type I'' theory with 
the shift.
First, let us notice that closed string states in 
(\ref{kamt}) with odd windings
enter only in the annulus amplitudes. This can be attributed to the
fact that only these states can be reflected by the standard O9-plane. 
The situation gets reversed if we now accompany the worldsheet  
parity operator with a $\sigma_{p_1}$ momentum shift along the circle.
This is done by replacing  the lattice sum $P_1(t)$
in the Klein bottle and Moebius strip  
amplitudes (\ref{akm}) by 
\be
P_1(t)\rightarrow P_1{0\brack {1\over 2}}(t) =\sum_{m_1\in {\bf Z}}\, 
(-)^{m_1} \,e^{-\pi t \alpha^\prime {m_1^2\over R_1^2}}
\label{p1-}
\ee
In the closed string channel this translates into the replacement
\be
W_1^{\rm even}({\ell\over 2})\rightarrow W_1^{\rm odd}({\ell\over 2})
\ee
and therefore now only odd windings modes are reflected by the
orientifold plane. We can see that combining this 
with a non-trivial Wilson line turned on in the D5 
gauge group, one gets the desired result. The Wilson line
can be included by replacing the lattice sum $P_1(t)$ accompanying
annulus terms linear in $k$ by
\be
P_1 (t)\rightarrow P_1{{1\over 2}\brack 0}(t) 
=\sum_{n_1\in {\bf Z}}\, 
e^{-\pi t \alpha^\prime {(n_1-{1\over 2})^2 \over R_1^2}}
\ee
After these replacements, the Klein bottle and 
annulus amplitudes can 
be written in the closed string channel as
\bea
\widetilde{\cal K}_0 &=&{2^5\over 2}\left[
 \chi_O+\chi_V+\chi_S+\chi_C\right]  
(i\ell)\, v_1 v_4 \tilde{v}_4\, W_1^{\rm odd}({\ell\over 2})\nonumber\\
\widetilde{\cal A}_0 &=&{2^{-5}\over 2}\left[ 
\chi_O I_O^2+\chi_V I_V^2 +\chi_S I_S^2+\chi_C I_C^2 \right] 
(i\ell)\, W^{\rm even}_1({\ell\over 2})\nonumber\\ 
&&+{2^{-5}\over 2}\left[
\chi_O I_S^2+\chi_V I_C^2 +\chi_S I_O^2+\chi_C I_V^2 \right] 
(i\ell)\, W^{\rm odd}_1({\ell\over 2})
\eea
The complete square is now reconstructed by  
\be
\widetilde{\cal M}_0 =-{2\over 2}\left[ 
\chi_O I_S+\chi_V I_C +\chi_S I_O+\chi_C I_V  \right]
(i\ell)\,\sqrt{v_1 v_4 \tilde{v}_4}\, W_1^{\rm odd}({\ell\over 2})
\ee      
which differ from the ones in (\ref{kamt}) by the 
parity of the closed string winding sum $W_1^{\rm odd}$ and in an overall
flip of the sign of $k$. This leads in the open string
channel to the lattice sum (\ref{p1-}) and the gauge group
$SO(M)\times SO(k)\times SO(N)$ gauge theory. In deriving
this result we have used the massless content 
$\rho_{00}=V+H$, $\rho_{0A}=V-H$ $\rho_{A0}={1\over 2}H$
with $V$, $H$ denoting massless $N=2$ vector- and hyper-multiplets.
For the case we are interested in,  we set $n_9=0$, which
leads, after the above replacements, to the direct amplitudes
\bea
{\cal K}&=&{1 \over 2}\,\rho_{00}(2it)P_1{0 \brack {1\over 2}}(t)
P_4(t) \tilde{P}_4(t)\nonumber\\ 
{\cal A}&=&{1 \over 2}\,\left[\rho_{00}\left({it\over 2}\right)\left(
n_5^2\,P_4(t) \tilde{W}_4(t)
+n_1^2\,W_4(t) \tilde{W}_4(t)\right)P_1(t)
\right.\nonumber\\&&\left.
+2 n_5 n_1 \rho_{A0}\left({it\over 2}\right)
\tilde{W}_4(t)P_1{{1\over 2}\brack 0}(t)\right]\label{akmf}\\
{\cal M}&=&-{1 \over 2}\,\left[
n_5\,\rho_{0B}\left({it\over 2}+{1\over 2}\right)P_4(t)
+n_1\,\rho_{0C}\left({it\over 2}+{1\over 2}\right)\right]
P_1{0\brack {1\over 2}}(t)
\eea  

\section{Appendix B: Symmetric product orbifold CFT:
free field theory realization}  
\renewcommand{\theequation}{B.\arabic{equation}}
\setcounter{equation}{0}

In this appendix we give an alternative derivation of
the partition  function formulae (\ref{11}) for the case where the  
Hilbert space ${\cal H}$ involved in the symmetric product
admits a free field theory (orbifold) description. We follow
closely \cite{bgmn}. 

The oscillator contribution of a given worldsheet field $\Phi$
with boundary conditions (\ref{bcphi}), to the string partition function 
is given by
\be
{\cal Z}_{\rm osc}{g_\phi\brack h_\phi}(q,y)=\prod_{n=1}^{\infty}
(1-e^{2 \pi i h_{\phi}} y^{\omega_\phi} q^{n-g_{\phi}})^{\epsilon_{\phi}}
\label{zp}
\ee
with $\epsilon_\phi=-1$ for bosons and $\epsilon_\phi=1$ 
for fermions. 
The partition function (\ref{zh}), can then be written as
a product over $\Phi$ of such contributions in the left
and right moving-part of the CFT, times
a (in general non-holomorphic) zero mode contribtuion   
\be
{\cal Z}{g\brack h}({\cal H}|q,\bar{q},y,\tilde{y})=\tau_2^{-{D\over 2}}
\prod_{\phi,\bar{\phi}} Z_0 Z_{\rm osc}\, {g_\phi\brack h_\phi}(q,y)
\bar{Z}_0 \bar{Z}_{\rm osc} \, 
{g_{\bar{\phi}}\brack h_{\bar{\phi}}}(\bar{q},\tilde{y})  
\ee
For complex bosonic and
fermionic degrees of freedom this zero mode contribution can 
be written as 
\bea
Z_0{0\brack 0}_{\rm boson}(q,y)&=& 
q^{\chi_\phi} q^{{1\over 2} P_L^2}\nonumber\\ 
Z_0{g_\phi\brack h_\phi}_{\rm boson}(q,y)&=&
q^{\chi_\phi} 
\nonumber\\
Z_0{0\brack 0}_{\rm fermi}(q,y)&=& q^{-\chi_\phi}
\,(y^{\omega_\phi}+y^{-\omega_\phi}-2)\nonumber\\
Z_0{g_\phi\brack h_\phi}_{\rm fermi}(q,y)&=&
q^{-\chi_\phi} 
\label{z0}
\eea
with similar expressions for the right-moving components in terms
of anti holomorphic quantities and $P_L$
replaced by $P_R$. In the following we will display only holomorphic
formulas since the analysis of the antiholomorphic part follows trivially. 
The boson in (\ref{z0}) is understood to be compact. For non compact
bosons we should of course simply omit the lattice sum in (\ref{z0}).  
$\omega_\phi$ stands for the charge of the field $\phi$ under $J_0^3$ and
$\chi_\phi$ represents the contribution of a complex boson with spin
characteristics $g_\phi,h_\phi$ to the zero point energy
\be
\chi_\phi=-{1 \over 12}+{1\over 2}g_\phi(1-g_\phi).
\label{energy0}
\ee

Orbifold group sectors and the $N$ copies of the field $\Phi$
are labeled following the notation of section 4. 
Since the $g$ and $h$ twists commute we can diagonalize them 
simultaneously. In this basis one can write 
\bea
{\bf g} &=& e^{2 \pi i\frac{l}{L}}\nonumber\\
{\bf h} &=& e^{2 \pi i(-{l s_{m;i}\over M L}+\frac{m}{M})}
\label{gh}
\eea 

Let us now evaluate the basic trace 
(\ref{bt}). For the time being 
we will concentrate on the left-moving contribution 
$Z_{\rm osc}{g_\phi \brack h_\phi}(q,y)$. After simple manipulations 
of the product formulae
one is left with
\bea
(M)^{r_M^L} {\bf t}\,
\Box \raisebox{-12pt}{\hspace*{-12pt} $(L)^{N_L}$}: 
&& 
\prod_{i,l,m}\prod_{n=1}^{\infty}
(1-e^{2\pi i ( -{l s_{m;i}\over M L}+{m\over M}+h_\phi)}\, y^{\omega_\phi}
q^{n-g_{\phi}-l/L})^{\epsilon_{\phi}}\nonumber\\
&&=\prod_i \prod_{n=1}^{\infty} 
(1-e^{2\pi i (s_i g_\phi+M h_\phi)}\, y^{M \omega_\phi}
(q^{M\over L}e^{2\pi i {s_{i}\over L}})^{n-g_{\phi}L})^{\epsilon_{\phi}}
\nonumber\\
&&=\prod_i 
Z_{\rm osc}{g_\phi L \brack g_\phi s_i+M h_\phi }(q^{M\over L}
e^{2\pi i {s_i\over L}} ,y^M)  
\eea
The result is in agreement with (\ref{bt}). 
One can follow similar manipulations to show that 
the zero mode contribution to $S_N {\cal H}$  
can be again reexpressed as
$Z_0{g_\phi L \brack g_\phi s_i^L+M h_\phi}
(q^{M\over L}e^{2\pi i {s_i^L \over L}}, y^M)$.
This is clear for the lattice sum and the fermionic zero mode
trace following similar manipulations as before,
while for the zero point energy this
can be read off from (\ref{energy0}) (let say in the 
$(L)^{M}$-twisted sector) leading to
a contribution $q^{\chi_{L,M}}$ with 
\bea
\chi_{L,M}&=& M \epsilon_\phi\, 
\sum_{l=0}^{L-1} \left[{1 \over 12}-{1\over 2}
(g_\phi+{l\over L})(1-g_\phi-{l\over L})\right]\nonumber\\
 &=&
  {M\over L}\epsilon_\phi\,\left[{1 \over 12}-{1\over 2}
  g_\phi L(1-g_\phi L)\right]
\eea
as expected. 

This concludes our derivation of (\ref{bt}) in the 
free field theory context.

Following the same steps as in section 4,  
one readily arrives
to the symmetric product formulae (\ref{11}).    

\section{ Appendix C: Modular integral with shifts}
\renewcommand{\theequation}{C.\arabic{equation}}
\setcounter{equation}{0}

In this appendix we evaluate the modular integral (\ref{g123}). 
\be
{\cal G}_{\vec{a},\vec{b}}(v,w)= 
{1\over 2}
\int {d^2 \tau\over \tau_2^{3+{d\over 2}}}\,\sum_{g,h,n}\, C{g\brack h} (n)
\Gamma_{2,2}{g\brack h}(v,w)\,e^{2\pi i n \tau}  
\label{intshift}
\ee
with $d=2(6)$ in the case of models $I_F$, $III_F$ ($II_F$) and
$\Gamma_{2,2}{g\brack h}(\ell\cdot\vec{v},\ell_*\cdot\vec{w})$ defined
by (\ref{latpert}).

We start by reabsorbing $(d+2)/2$ powers (the number of $\vec{w}$-insertions)
of $\tau_2$ by the rescaling 
of the left moving source $\vec{w}\rightarrow {\vec{w}\over \tau_2}$.
Equalities in the following  are understood to hold 
once $(d+2)/2$ $\vec{w}$-derivatives      
are applied to the final result, with the sources $\vec{w}$ 
then set to zero.

After performing the $\tau_1$ gaussian integral we are left with
\be
{\cal G}_{\vec{a},\vec{b}}={\cal C}\,{(U_2 T_2)^{1\over 2}\over |m_1|}
\int_0^\infty  {d \tau_2\over \tau^{3/ 2}_2}\, e^{-b_0-\gamma \tau_2-
{\beta\over \tau_2}}={\cal C}\,{(U_2 T_2)^{1\over 2}\over |m_1|}
\sqrt{\pi\over \beta}\, e^{-b_0-2\sqrt{\beta\gamma}}
\label{intt2}
\ee
with
\bea
{\cal C}&=&{1\over 2}\sum_M \eta{g\brack h} C{g\brack h} (n,{\bf l})\nonumber\\
b_0&=&-2\pi i T_1 m_1 n_2-2 \pi i {n\over m_1}(n_2 U_1 +n_1)\nonumber\\
&&+2\pi i \ell\cdot \left[ \vec{n}\vec{v}-v_1(n_2 U_1+n_1)\right]
-2\pi i\ell_*\cdot w_1\left[m_1- {U_2\over m_1 T_2}(n+m_1 \ell\cdot v_1)\right]
\nonumber\\
\beta&=&\pi n_2^2 T_2 U_2+2\pi i n_2 \ell_*\cdot(w_1 U_1-w_2)+
\frac{\pi U_2}{T_2}(\ell_*\cdot w_1)^2\nonumber\\
\gamma&=&m_1^2{\pi T_2\over U_2}\left[1+{U_2\over m_1^2 T_2}(n+m_1\ell\cdot v)
\right]^2
\eea
We are interested in the leading order in a $1/T_2$ 
expansion of (\ref{intt2}), which is
associated in the dual theory to the semiclassical 
approximation around the D-instanton background. 
In this limit all subleading $\vec{w}$-dependent terms in 
the exponential can be discarded since they lead to subleading
contributions once they are hitted by $\vec{w}$-derivatives.
At the leading order the result can be written as
\be
{\cal G}_{\vec{a},\vec{b}}=\sum_{m_1,n_2,g} {1\over |n_2|}e^{2\pi i \left[
T m_1 n_2 +{n\over m_1}n_2U+n_2 \ell\cdot \hat{v}-m_1\ell_*\cdot \hat{w}\right]}
\, {\cal J}_{\vec{a},\vec{b}}+h.c.  
\label{final}
\ee
where $\hat{v}=v_1 U-v_2$ and $\hat{w}={1\over U_2}(w_1 \bar{U}-w_2)$ 
are induced sources
and $h.c.$ denotes the hermitian conjugate contributions coming from
anti D-instantons.
${\cal J}_{\vec{a},\vec{b}}$ represents the shift dependent
sum
\be
{\cal J}_{\vec{a},\vec{b}}={1\over 2 |m_1|}\sum_{n_1,g,h}\, 
e^{2\pi i ({n\over m_1}n_1-2 a_1 b_1 gh-a_1 h m_1+a_1 g n_1)}
C{g\brack h}(n,{\bf l}).
\label{p}
\ee
Our next task is to evaluate (\ref{p}) in the cases of a winding 
$a_1=1$ or
momentum shift $b_1=1$. The domains of $\vec{m},\vec{n}$ are
specified by (\ref{M}), while $n$ is integer in the untwisted 
sector and both integer and half integer in the twisted sector.
Using the identity
\be
C{{1\over 2}\brack {1\over 2}}(n,{\bf l})=(-)^{2n}
\,C{{1\over 2}\brack 0}(n,{\bf l})
\ee
and performing the geometric sums one can write the final 
result as 
\bea
{\cal J}_{1,0}&=& 
{1\over 2}(C{{1\over 2}\brack{1\over 2}} +(-)^{m_1}
C{{1\over 2}\brack 0})(n,{\bf l})\,
{\cal P}_{n\over m_1}^{\bf Z} 
+{1\over 2}
C{ 0\brack{1\over 2}}(n,{\bf l}) \,{\cal P}_{n\over m_1}^{\bf Z+{1\over 2}}
\nonumber\\ 
{\cal J}_{0,1} &=& 
{1\over 2}(C{{1\over 2}\brack{1\over 2}}
+(-)^{n\over m_1}C{{1\over 2}\brack 0}
+2C{ 0\brack{1\over 2}})(n,{\bf l})\,
{\cal P}_{n\over m_1}^{\bf Z}  
\eea
with ${\cal P}_{n\over m_1}^{{\bf Z}+\delta}$ a projector onto
states with ${n\over m_1} \in {\bf Z}+\delta$. 

Plugging in (\ref{final}) and introducing the quantum number 
$k\equiv {n\over m_1}\in {\bf Z}+a_1 g$ one obtains,
after performing the remaining $m_1,n_2$ sums, with 
the final result (\ref{genusf}).

\rnc{\Large}{\normalsize}

\end{document}